\documentstyle[12pt]{article}

\date{March 11, 1999}

\def\be{\begin{equation}}
\def\ee{\end{equation}}
\def\bear{\begin{eqnarray}}
\def\eear{\end{eqnarray}}
\def\nn{\nonumber}

\def\half{{{1\over 2}}}

\def\Im{{\rm Im\hskip0.1em}}

\def\Id{{\bf I}}                                

\newcommand\px[1]{{\partial_{#1}}}

\newcommand\rep[1]{{\underline{\bf {#1}}}}      
\newcommand\tr[1]{{\mbox{tr}\{{#1}\}}}          
\newcommand\com[2]{{\lbrack {#1},{#2}\rbrack}}  

\def\BZ{{{\bf Z}}}
\def\BC{{{\bf C}}}
\def\BR{{{\bf R}}}
\newcommand\MR[1]{{{\bf R}^{#1}}}               
\newcommand\MC[1]{{{\bf C}^{#1}}}               
\newcommand\MS[1]{{{\bf S}^{#1}}}               
\newcommand\MT[1]{{{\bf T}^{#1}}}               

\newcommand\SUSY[1]{{{\cal N}= {#1}}}           

\def\a{{\alpha}}
\def\b{{\beta}}
\def\g{{\gamma}}

\def\u{{\mu}}

\def\tht{{\theta}}

\def\lam{{\lambda}}

\def\ZM{{\Xi}}   
\def\MTW{{\widetilde{{\cal M}}}}   
\def\CTO{{\cal T}} 
\def\Abl{{\bf A}} 
\def\Nil{{\bf N}} 
\def\Kom{{\bf K}} 
\def\Grp{{\bf G}} 

\def\calh{{\cal H}} 
\def\calg{{\cal G}} 
\def\Weyl{{\cal W}} 

\def\vt{{\theta}}  
\def\BTW{{\widetilde{{\cal B}}}}  

\newcommand\lrp[2]{{{\langle {#1}, {#2} \rangle}}} 

\newcommand\emph[1]{{\em {#1}}}  

\begin{document}

\begin{titlepage}
\titlepage
\rightline{hep-th/9903110, PUPT-1836}
\rightline{March 11, 1999}
\vskip 1cm
\centerline{{\Large \bf Two Conjectures on Gauge Theories, Gravity,}}
\centerline{{\Large \bf and Infinite Dimensional Kac-Moody Groups}}
\vskip 1cm
\centerline{
Ori J. Ganor\footnote{origa@puhep1.princeton.edu}
}
\vskip 0.5cm

\begin{center}
\em  Department of Physics, Jadwin Hall \\
Princeton University\\
NJ 08544, USA
\end{center}

\vskip 0.5cm

\abstract{
We propose that the structure of gauge theories, the $(2,0)$ and
little-string theories is encoded in a unique
function on the real group manifold $E_{10}(\BR)$.
The function is invariant under the maximal compact subgroup $K$
acting on the right and under the discrete U-duality subgroup
$E_{10}(\BZ)$ on the left. 
The manifold $E_{10}(\BZ)\backslash E_{10}(\BR) /K$ contains
an infinite number of periodic variables.
The partition function of $U(N)$, ${\cal N}=4$
Super-Yang-Mills theory on $T^4$, with generic $SO(6)$ R-symmetry twists,
for example,
is derived from the $N^{th}$ coefficient of the Fourier transform
of the function with respect to appropriate periodic variables,
setting other variables to the R-symmetry twists and the
radii of $T^4$.
In particular, the partition function of nonsupersymmetric
Yang-Mills theory is a special case, obtained from the twisted $(2,0)$
or little-string theories.
The function also seems to encode the answer to questions about M-theory
on an arbitrary $T^8$.
The second conjecture that we wish to propose is that
this function is harmonic with respect
to the $E_{10}(\BR)$ invariant metric.
In a similar fashion, we propose 
that there exists a function on 
the infinite Kac-Moody group $DE_{18}$ that encodes
the twisted partition functions of the $E_8$ 5+1D theories as well as
answers to questions about the heterotic string on $T^7$.
}
\end{titlepage}


\tableofcontents

\section{Introduction}
Over the past few years there has been an accumulation of discoveries
indicating that gravity and gauge fields are different facets
of the same theory.
On the one hand, gauge theories are realized \cite{WitBND} as low-energy
descriptions of D-brane \cite{PolDBR} dynamics.
On the other hand, the S-matrix of  certain flat vacua of M-theory
can be mapped to questions about the large $N$ limit of gauge theories
\cite{BFSS,Juan,GKP,WitAdS}
(and see \cite{GubKle} for preliminary results).

In this paper we are going to propose a construction for a generating
function for the partition functions of Yang-Mills theories as
well as higher dimensional generalizations on tori.
 Let $Z_N$ be
the partition function of a $U(N)$ Yang-Mills theory on $\MT{4}$
(ignoring for the moment details like electric/magnetic fluxes and twists).
Also, let $Z_0$ be a quantity that will be specified later.
By ``a generating function'' we mean,
\be\label{zphi}
Z(\phi) \equiv \sum_{N=0}^\infty e^{-N t} Z_N \cos N\phi.
\ee
Here $\phi$ is periodic with period $2\pi$ and $t$ is a positive
real number that was added for convergence.
This partition function is a special limit of
a more general question, namely the partition function of the $(2,0)$
theory on $\MT{6}$, and the latter is a special case of an even
more general question about the partition function of the 6D
little-string theories \cite{SeiVBR} with $\SUSY{(1,1)}$ or
$\SUSY{(2,0)}$ supersymmetry.
These theories have a positive integer parameter $N$, like $U(N)$ 
Yang-Mills theories, that can be used to construct the generating
function.
As it stands, (\ref{zphi}) is ill-defined for the $(2,0)$ and
little-string theories because
these theories have a noncompact moduli space of vacua 
and without further modifications the integrals that define
the partition functions on $\MT{6}$ are divergent.
We will therefore modify the definition of the partition function
by adding R-symmetry twists.
The generating function for partition functions (\ref{zphi})
now depends on the shape and size of the $\MT{6}$, the values of the
twists and the phase $\phi$.
We will argue that these parameters can be naturally embedded inside
an infinite dimensional manifold,
$$
\MTW \equiv E_{10}(\BZ)\backslash E_{10}(\BR)/\Kom,
$$
where $E_{10}(\BR)$ is the infinite dimensional
group constructed from the infinite dimensional Generalized
Kac-Moody algebra $E_{10}$.\footnote{We will henceforth refer to
Generalized Kac-Moody algebras simply as ``Kac-Moody'' algebras.
What is sometimes referred to in physics
as ``Kac-Moody'' algebras is really an ``affine Lie algebra'' which
is a special case of Kac-Moody algebras.}
where $\Kom$ is a maximal compact subgroup.
We propose that there exists a {\em Master-Function}
$\ZM(\xi)$ defined for $\xi\in\MTW$ which encodes, among other things,
the partition functions of Yang-Mills theories.
We will also argue that the ``quiver'' theories of \cite{DouMoo}
can be read off from such a $\ZM$.

We propose to construct $\ZM$ as follows.
Start with M-theory on $\MT{9}$. The transverse space has
an $SO(2)$ rotation symmetry and we can define 9 $SO(2)$ twists
along the cycles of $\MT{9}$. We propose that for generic irrational
twists the partition function of M-theory on such a space is well defined
and not identically equal to 1. The partition function depends on
the size and shape of the $\MT{9}$ as well as on various fluxes
along it. It also depends on the $SO(2)$ twists.
We will argue that the parameter-space of the
geometrical $SO(2)$-twists should be augmented with (formally) all
of the U-dual twists and that together they form the manifold $\MTW$.
The partition function of the little-string theory at rank $N$
can be read off from $\log(\ZM)\,$ by (slightly oversimplifying)
extracting the contribution of 
an instanton made by wrapping $N$ M5-branes along $\MT{9}$
and by taking the limit that all the radii except 7 become infinite.
These 7 radii correspond
to the size of $\MT{6}$ on which the M5-branes wrap and the size
of another circle that will play the role of the type-IIA dilaton
for the construction of the little-string theory.
It will actually turn out that in order to incorporate the twists we need
to immerse the $N$ M5-branes inside an instanton made by wrapping
a Kaluza-Klein monopole around $\MT{6}$ and further immerse all of that
inside an instanton that can be defined algebraically
by formally wrapping a D8-brane of type-IIA.
One can isolate the contribution of this configuration because
each instanton couples to a flux of its own that is a periodic variable
inside $\MTW$ and can be Fourier transformed.

We can now reveal the (conjectured) identity of $Z_0$ in (\ref{zphi}),
or rather its analog for the little-string theories.
We propose that it is the ``gravity-part'' of the partition function
of M-theory on $\MT{7}$ with twists in the $Spin(4)$ which rotates
the transverse space. By ``gravity-part'' we mean the contribution
without the M5-brane instantons.
It therefore seems that the Master-Function $\ZM$ also encodes
information about the S-matrix of massless particles
in M-theory on $\MT{8}$.

Finally, based on the leading behavior of the instanton terms,
we will conjecture that $\ZM$ is a harmonic function on $\MTW$
with respect to the $E_{10}(\BR)$ invariant metric and that
for the appropriate boundary conditions, namely to vanish at
the various decompactification limits, it is unique.

The paper is organized as follows.
In section (2) we discuss the twisted field-theory partition functions.
In section (3) we introduce the partition functions of M-theory on
a twisted space.
In section (4) we review some basic facts about finite
Kac-Moody algebras and their related groups.
In section (5) we will review the relation between
the positive roots of $E_8$ 
and the instantons of M-theory on $\MT{8}$.
In section (6) we review some basic facts about infinite
Kac-Moody algebras. In section (7) we discuss the relevant properties
of $E_9$ and $E_{10}$ in relation to M-theory.
In section (8) we present the algebraic formulation of the
conjecture and in section (9) we conclude with a discussion and suggestions
for extensions of the conjecture.

\subsection*{A bibliographical note}
The group $E_{10}$ was introduced in \cite{JulEX}
as a generalization
of the moduli spaces of supergravity to lower dimensions.
Evidence for this suggestion was given in \cite{NicEX}.
The T-duality group in complete compactification of string-theory
with a compact time-like direction was studied in \cite{MooFIN}.
The discrete subgroup $E_{10}(\BZ)$ was proposed as a U-duality
group of M-theory on $\MT{10}$ in \cite{HulTow}.
The maximal compact subgroup of $E_{10}(\BR)$ was described in
\cite{NicII}. In that paper the coset space
$E_{10}(\BZ)\backslash E_{10}(\BR)/\Kom$ is also discussed, though
the interpretation is different than ours.
A progress in understanding the representation of $E_{10}$ was made
in \cite{GebNicI,GebNicII}.
The relation between $E_{10}$ and M-theory has also been studied in
\cite{Mizog,MartEX,GebMiz}.

In a different context, $E_{10}$ has recently discussed as a classifying
symmetry for the spectrum of compactified 6D $(1,0)$ theories
\cite{DHIZi,DHIZii}.
Also, in \cite{DVVKM} another infinite dimensional Kac-Moody algebra
appeared in the context of counting BPS states.

Recently, somewhat similar generating functionals for Chern-Simons
theory have been developed in \cite{GopVafI}-\cite{GopVafIV}.
Also recently, lower dimensional compactifications of M-theory,
their limits and their U-duality groups have been studied in \cite{BFM}.

The study of instanton terms in M-theory is an active area of research
\cite{GGV}-\cite{OPNew}.
The harmonic equation for the instanton terms has been proposed
in \cite{GGV} for 9+1D type-IIB and later generalized to lower dimensions.
It has been proved in 9+1D in \cite{GreSet} using supersymmetry
(see also \cite{PiolH}).

\section{The twisted ``field-theory'' partition functions}
In principle, one can recover the spectrum and S-matrix of a field-theory
 from the partition function of the theory on $\MT{4}$.
Let this partition function be $Z(R_0,R_1,R_2,R_3)$ where $R_i$ are
the radii of the cycles and we assume that $\MT{4}$ is of the form,
$\MS{1}\times\cdots\times\MS{1}$.
Non-supersymmetric Yang-Mills theory, for example,
would have a discrete spectrum
on $\MT{3}$ and by expanding $Z$ asymptotically as $R_0\rightarrow\infty$
one can read-off the energy spectrum of QCD on $\MT{3}$ (in a given
sector of fluxes).  In the limit, $R_i\rightarrow\infty$,
one should be able to recognize the spectrum as the free Fock space of
the glueballs. The splitting between energy levels is of the order
of $R_i^{-1}$ and the corrections to the energy of, say, a 2-particle
state is proportional to $V^{-1}$ where $V\equiv R_1 R_2 R_3$.
As $R_i\rightarrow\infty$ the $V^{-1}$ corrections
are much smaller and therefore the S-matrix can also be read-off from
these corrections and hence from $Z$.

The theories that we are interested in require a slight modification.
Let us take for example $U(N)$ SYM with $\SUSY{4}$ supersymmetry.
The partition function on $\MT{4}$ is not a good object because:

\begin{itemize}
\item
After compactification on $\MT{3}$,
the moduli-space $\MR{6N}/S_N$ of the 3+1D theory would give rise to 
$6N$ non-compact integration variables.
\item
Even if we worked with $\SUSY{1}$ SYM, which has no moduli, the partition
function would still not be useful because it will just equal $N$ --
the Witten index \cite{WitIND}.
\end{itemize}

The solution to the problem is to impose twisted boundary conditions
(see also the related papers
\cite{WitIND}-\cite{CGKM}).
The $\SUSY{4}$ theory has an $SU(4)$ R-symmetry.
We can now define,
$$
Z_N(R_0,\dots R_3; g_0, \dots g_3),
\qquad g_i\in SU(4),\qquad i=0\dots 3,
$$
to be the partition function of $U(N)$ $\SUSY{4}$ SYM on
$\MS{1}\times\MS{1}\times\MS{1}\times\MS{1}$ such that the
periodicity conditions on the fields are given by $g_0,\dots, g_3$.
This means that as we go around the $i^{th}$ $\MS{1}$ the fields 
are rotated by $g_i\in SU(4)$. The requirement on the $g_i$'s is that
they should be mutually commuting.
Hence, they can be taken to be inside a Cartan torus
$g_i\in\CTO\equiv\MR{3}/\Gamma(A_3)$,
where $\Gamma(A_3)$ is the root lattice
of $SU(4)$. On top of that, we have to mod out by
the Weyl group $\Weyl=S_4\subset SU(4)$.
Let us also take a generic $\MT{4}$
that is parameterized by a point in,
$$
SL(4,\BZ)\backslash GL(4,\BR)/SO(4).
$$
Thus, the parameter space for $Z_N$ is,
$$
\MTW_{SU} \equiv
SL(4,\BZ)\backslash ((\CTO^4/S_4)\times GL(4,\BR))/SO(4).
$$
Here $h\in SL(4,\BZ)$ acts on $g\in GL(4,\BR)$ as $g\rightarrow h\circ g$
and on $\CTO^4$ it acts in the fundamental representation $\rep{4}$,
with $\CTO$ treated as a $\BZ$-module. The Weyl group $S_4$  acts
on all four $\CTO$ factors simultaneously.

 For generic twists, the moduli fields are no longer zero modes and
supersymmetry is broken. Therefore, $Z_N(\xi)$, for $\xi\in\MTW_{SU}$,
is a non-trivial function.

Similarly, we can define the partition function of the $(2,0)$
theories on $\MT{5,1}$ with $Spin(5)$ twists. In \cite{WitAdSII},
it has been argued that the $(2,0)$ theory compactified
on $\MT{2}$, with a twist that is the element $(-1)\in Spin(5)$ along
the large cycle of $\MT{2}$,
approaches ordinary non-supersymmetric QCD
when the complex structure, $\tau$, of $\MT{2}$ is taken to infinity.
Thus, the analogous function for the $(2,0)$ theory encodes
the spectrum of ordinary QCD.

We can even go up one step to the little-string theories \cite{SeiVBR}.
In light of \cite{MalStr}, the spectrum of the little-string theories
on $\MT{5}$ is probably not discrete (see also \cite{GubKle,ABKS}
for related issues),
it is likely that the twisted partition functions $Z_N$ 
are well-defined -- they just do not have a discrete Fourier
expansion in time.

The twisting of the little-string theories actually involves
more parameters than just the geometrical ones.
In \cite{CGK}, it has been argued that because the little-string
theories have the T-duality property, one should also consider the
T-duals of the twists.
The twists are commuting elements of $Spin(4) = SU(2)\times SU(2)$.
Altogether, the twists and their T-duals form the representation
$\rep{2d}$ of the duality group $SO(d,d,\BZ)$.
The cartan torus is now simply $\MS{1}\times\MS{1}$ and the Weyl group is
$\BZ_2\times\BZ_2$.
Thus, it is natural to conjecture that
the full parameter-space is,
$$
\MTW_{LS} \equiv
SO(6,6,\BZ)\backslash (((\MS{1}\times\MS{1})^{12}/(\BZ_2\times\BZ_2))
\times (SO(6,6,\BR)/SO(5,1))).
$$
We have a partition function $Z_N(\xi)$
for $\xi\in\MTW_{LS}$ and $N\in\BZ_{+}$.
We have seen that expanding it in special asymptotic limits in $\MTW_{LS}$
reproduces the spectrum of QCD and of many other theories.

In the next section we will argue that all the $Z_N$ functions
can be encoded in special asymptotic limits of a single ``Master-Function''
$\ZM$ and this function also encodes gravity.


\section{Twisted toroidal M-theory partition functions}
We will now propose a construction of M-theory partition functions
using a similar idea as in the previous section.
 We will later argue that these functions are the generating
functions of the field-theory partition functions.

\subsection{Twisting M-theory}
Let us start with M-theory on $\MT{6,1}\times\MR{4}$ and add a twist.
Thus, we choose 7 commuting elements $g_i$ of $Spin(4) = SU(2)\times SU(2)$
and postulate that as we go around one of the cycles of $\MT{6,1}$
we have to rotate $\MR{4}$ by the appropriate group element.
We have thus replaced $\MT{6,1}\times\MR{4}$ with an $\MR{4}$ fibration
over $\MT{6,1}$. Let us denote this space by $Y$.
For generic twists, $g_i$, the supersymmetry is completely broken.
We would like to propose that there exists a well-defined partition
function $\ZM_7$ of M-theory on $Y$ which is a function of the $g_i$'s and
of the moduli of $\MT{6,1}$. We will describe in more detail
what $\ZM_7$ is a function of later on.

The heuristic argument for the existence of the partition function
goes as follows.
The space $Y$ is unbounded so in principle the partition function
could be infinite. However, this particular space $Y$ behaves in
a special way at infinity. Let us take the Euclidean version of $Y$,
let $x$ be a coordinate on $\MT{7}$ (instead of $\MT{6,1}$)
and let $y$ be a coordinate on the $\MR{4}$-fiber.
Let us also assume that none of the twists $g_i$ is ``rational''
(i.e. there is no integer $n$ such that $g_i^n = 1$).
Then, given any arbitrarily large $\Lambda$ we can find a $\Lambda'$
such that any point with $\| y\| > \Lambda'$ has an 11-dimensional
ball $B^{11}$ of radius $\Lambda$ that surrounds it and such
that the metric in $B^{11}$ is flat and no two points inside
$B^{11}$ are identified. If there are $k$ rational twists, the
statement has to be modified by replacing $B^{11}$ with
$B^{11-k}\times\MT{k}$ where $\MT{k}$ is of fixed size and shape.
In any case the point is that, asymptotically, the space $Y$ looks
locally like flat $\MR{11-k}\times\MT{k}$.
Because of supersymmetry, the action of M-theory on $\MR{11-k}\times\MT{k}$
is identically zero. Thus, the main contribution to the action of M-theory
on $Y$ is localized in the region around
the origin $y=0$ of the fiber $\MR{4}$.
The contribution of points with very large values of $\| y\|$ can only
come from non-local effects like large loops of particles and
we assume that these effects are negligible as $\| y\|\rightarrow\infty$.

\subsection{A closer look at the twists}
What is the parameter-space on which $\ZM_7$ is defined?
As defined, $\ZM_7$ is a function of the moduli,
$$
g\in E_7(\BZ)\backslash E_7(\BR)/SU(8),
$$
of the $\MT{7}$ and of the 7 twists, $g_0\dots g_6$, which
can be taken in the toral subgroup $U(1)\times U(1)\subset Spin(4)$
subject to the overall $\BZ_2\times\BZ_2$ Weyl-group identification.
However, if we act with an element of the U-duality group
$E_7(\BZ)$, a geometrical twist does not necessarily transform
into a geometrical twist. Thus, just like the $\eta$-twists of the
little-string theory (see \cite{CGK}), the parameter-space
of the twists is actually larger.
To see what it is, let us assume that all the twists are ``rational''
(i.e. that there is an integer $n$ such that $g_i^n = \Id$).
In this case we can think of the space $Y$ as obtained by modding out
the background $\widetilde{\MT{7}}\times\MR{4}$, by a 
finite abelian group $A$. Here $\widetilde{\MT{7}}$ is a torus
which is $n$ times larger than the original $\MT{7}$ in all directions.
$A$ is generated by 7 generators $g'_i$, where $g'_i$ acts simultaneously
as $g_i$ on the transverse space $\MR{4}$ and as a shift
in the $i^{th}$ direction of $\widetilde{\MT{7}}$.
Another way of saying the same thing is that in the original $\MT{7}$,
a state with some $U(1)\times U(1)\subset Spin(4)$
charge (rotations in the transverse $\MR{4}$), has fractional momentum
in $\MT{7}$ and the twists relate the non-integer part of the momentum
to the $Spin(4)$ charge.

The 7 components of the momentum in $\MT{7}$ are part of an
$E_7(\BZ)$ multiplet of $\rep{56}$ charges.
Thus, we see that the natural way to make the parameter-space U-duality
invariant is to postulate a $U(1)\times U(1)\subset Spin(4)$ twist for 
every one of the $56$ charges.
We will assume (without proof) that all these twists have to be commuting
and that the $\BZ_2\times\BZ_2$ Weyl group acts simultaneously on all
of them. Thus, the parameter space is,
\be\label{pmspsev}
\MTW = 
E_7(\BZ)\backslash
\left(
 \left((U(1)\times U(1))^{56}/(\BZ_2\times\BZ_2)\right)\times
 \left(E_7(\BR)/SU(8)\right)
\right).
\ee
Here, $E_7(\BZ)$ acts on all the factors simultaneously.

In addition to the 7 geometrical twists
there are now 21 twists corresponding to fractional M2-brane charges,
21 twists corresponding to fractional M5-brane charges and
 7 twists corresponding to fractional KK-monopole charges.
The extra non-geometrical
twists are defined as U-duals of the geometrical twists.
For example a twist that corresponds to fractional M2-brane charge
in directions $1,2$ can be defined as the limit of M-theory
on $\MT{3}\times\MT{4}$ with $\MT{4}$ large and $\MT{3}$ very small
and with a geometrical twist in the $3^{rd}$ direction.
An interesting question might be whether these twists have
another low-energy description without using U-duality.
The twisting by
the KK-monopole charge seems to be a particularly interesting case
because, for large radii, it might be possible to describe it
with some unconventional geometry.
For large radii, a KK-monopole can be described as 
a smooth geometrical solution.
The KK-monopole charge with respect to the, say, $7^{th}$ direction
in $\MT{7}$ corresponds to the first Chern class of
the circle bundle that the $\MS{1}$ (in this $7^{th}$ direction)
makes over an $\MS{2}$ that surrounds the KK-monopole at infinity.
A fractional KK-monopole charge means that the first Chern class
is fractional if it comes with a state that is in a non-trivial
representation of the transverse $Spin(4)$.
It is interesting to note that fractional Chern classes naturally
arise in the context of noncommutative geometry \cite{CDS,DH}
and it might have something to do with this kind of twist.

\subsection{Instanton expansion}
We will now argue that $\ZM_7$, as defined above, encodes all the
partition functions $Z_N$ of the twisted little-string theories
(and hence also ordinary non-supersymmetric Yang-Mills theories).

To see this, let us take $\MT{7}$ in the form $\MT{6}\times\MS{1}$
and let us, as a preliminary step,
shrink $\MS{1}$ to zero size, keeping $\MT{6}$ finite
in string units just as in \cite{SeiVBR}. Let us also set the twist
along $\MS{1}$ to zero. We obtain weakly coupled type-IIA on
$\MT{6}$ with transverse-$Spin(4)$ twists.
However, $\ZM_7$ also depends on the 6-form (dual to the NSNS 2-form)
flux along $\MT{6}$. This flux is a periodic variable
in (\ref{pmspsev}) (at least if we keep the $Spin(4)$-twists to be
geometrical or take $SO(6,6,\BZ)$ T-dual $\eta$-twists).
Let us call it $\phi$.

What is the Fourier transform of $\log(\ZM_7)\,$ with respect to $\phi$?
Imagine an instanton made out of $N$ NS5-branes wrapped on $\MT{6}$.
If we assume that the twists are irrational, then the NS5-branes
are pinned to the origin of the transverse $\MR{4}$.
(If,  to the contrary,
there exists an $n\in\BZ_{+}$ such that $g_i^n = \Id$ for all $i=0\dots 6$
then a configuration of $n$ NS5-branes can separate to infinity.)
Such an instanton will be accompanied by a factor \cite{BBS,WitNON},
$$
Z_N e^{-N({V\over {\lam^2}} + i\phi)},
$$
where $\lam$ is the string coupling constant and $V$ is the volume of
$\MT{6}$, in string units.
The prefactor $Z_N$ is precisely the partition function
of the twisted little-string theory.
There is a subtlety about the treatment of electric and
magnetic fluxes, but that will be discussed later on.
 For the time being we note that the type-IIB little-string
theory has a low-energy description of 5+1D $U(N)$ SYM and hence
sectors with specified electric and magnetic fluxes.
The type-IIA little-string theory also has tensor fluxes \cite{GanSet}
and a detailed description of the different sectors is given in 
\cite{AhaWit,WitFLX}.  The electric and magnetic sectors can be
read off from $\ZM_7$ if we Fourier expand in the NSNS 2-form fluxes
along $\MT{6}$. The reason is that an electric/magnetic flux along a face
of $\MT{6}$ couples to the  appropriate 2-form flux.
We will return to that point in subsection~(\ref{subsec:emflx}).

Similarly,  by taking other limits of the parameter space and expanding
in different periodic variables one can recover the partition functions of
other branes. (Of course, one can recover them from the $Z_N$ as well.)

\subsection{Possible singularities of $\ZM_7$}\label{erratic}
The above discussion might have given the misleading impression
that $\ZM_7$ is a smooth function.
We will now argue, to the contrary, that $\ZM_7$ in fact exhibits
a very pathological behavior.
Let us first discuss the possible singularities
of $\ZM_7$. 
Our central assumption above was that the action of M-theory
on $Y$ is concentrated on a region of finite volume.
This assumption fails in the following case.
Suppose all the twists $g_i$ preserve a vector in the representation
$\rep{4}$ of $Spin(4)$. This means that there is a direction
in the transverse space, $\MR{4}$, which does not participate
in the twisting. Thus, translational invariance along this direction
is preserved and the partition function will be multiplied by
the infinite length of this direction.

There is another situation which is potentially dangerous too.
Suppose all the twists $g_i$ are rational but nonzero.
There is an integer $n$ such that $g_i^n=\Id$ for $i=0\dots 6$.
Let $H$ be the subgroup of $Spin(4)$ generated by
all the $g_i$'s and let $m={\mbox{\rm ord}}\,H$,
the number of elements in $H$.
The problem arises because an instanton made out of $m$  branes wrapping
some cycle of $\MT{7}$ can separate into $m$ branes which occupy an
orbit of $H$ in $\MR{4}$. Because the separation mode is noncompact
there is the potential danger that the contribution of this
particular  instanton will diverge.
However, it seems that this problem is actually averted as follows.
When the branes are separated by a large distance in 
the transverse $\MR{4}$ we can neglect the modes coming from strings
or membranes connecting different branes. The $m$ branes actually 
form one large connected brane in the covering space.
In any case, the brane which is far away on the transverse $\MR{4}$
is described, to a good approximation, by a theory with 16 supersymmetries
because the twist does not have a fixed point at the position of the
brane anymore. Thus, the contribution to the instanton from these
separated configurations is likely to converge.

It is intriguing to note the resemblance with a situation that
occurs frequently in non-commutative geometry.
Take a non-commutative gauge theory on $\MT{2}$ for example.
For rational non-commutativity parameters it can be mapped
to a commutative gauge theory on a smaller $\MT{2}$ but with a larger
gauge group and with some magnetic flux (see \cite{CDS,DH}).
On the other hand, the rational points can always be described
as limits of the irrational points. (See also \cite{Kol}.)

Finally, there is a case which seems certain to lead to a singularity
of $\ZM_7$.
This is when one of the twists is the element $(-1)\in Spin(4)$.
In this case, the twist is by the fermion number $(-)^F$. For a Euclidean
space-time, such a twist is associated with having a finite temperature.
Even if the other twists are irrational, the partition function is
very likely to diverge.
This problem occurs whenever the twist is of the form
 $e^{{{2j+1}\over {2 k}}\pi i}$, for $j,k\in\BZ$.
For a recent discussion on the behavior of string theories in
finite temperature see \cite{ABKR}.

It might seem that because of the erratic behavior of $\ZM_7$, it is
ill-defined.
To argue to the contrary, let us give an example.
Take,
\be\label{fqz}
f(q,z=r e^{i\phi})\equiv \sum_{N=1}^\infty {{z^N}\over {1 - q^N}},\qquad
|z| < 1.
\ee
This function has a pole in $q$, for $q=e^{{l\over k}\pi i}$ (and
integer $k,l$). These points are dense on the circle $|q|=1$.
On the other-hand, if $q=e^{\lam\pi i}$ and $\lam$ is an irrational
number, the function can be well defined.
For example, for $\lam=\sqrt{2}$, one can easily show that for all
$k,l\in\BZ_{+}$,
$$
|\sqrt{2} - {l\over k}| > {1\over {2\sqrt{2} k^2}},
$$
and therefore (\ref{fqz}) converges for $|z|< 1$.
A similar argument holds for any algebraic (i.e. a root
of a polynomial with integer coefficients) irrational $\lam$.
Furthermore,
if we Fourier decompose $f(q,z)$ for fixed $|z|$ and with respect
to $\phi = {\mbox{arg}} z$, then for fixed $N$ we get the function
$1/(1-q^N)$ which is smooth except for $N$ singularities.
Also, if we extend the definition of $f(q,z)$ for $|q|<1$ we get
a well-behaved function.

What is the moral of the story? If $q$ is the analog of the twist
and ${\mbox{arg}} z$ is the analog of the phase that couples to the
NS5-brane then,

\begin{itemize}
\item
It is plausible that $\ZM_7$ is well defined for all irrational
twists and that its Fourier transform that extracts
$Z_N$ makes sense.

\item
It is plausible that by extending the definition of $\ZM$, by finding
an extra parameter that is the analog of $|q|$, we get a smooth
function. Below we will propose that this parameter
completes the domain of the function to a coset of the $E_{10}$ group
manifold.

\end{itemize}


\subsection{Higher dimensional tori}\label{subsec:zmhdt}
Can we extend this construction to $\MT{d}$ with $d>7$?
In particular, can we trade some of the twists with moduli of $\MT{d}$?

Let us start with the case $d=8$. The transverse space is $\MR{3}$
and all the twists are in $Spin(3) = SU(2)$.
However, in this case we face one of the problems discussed in the
previous subsection -- a problem that
arises whenever $(11-d)$ is odd.
Commuting elements of $Spin(3)$ always have a common eigenvector
in the representation $\rep{3}$.
This means that there will always be a direction
in the transverse $\MR{3}$ which is preserved by all the twists and
hence translational invariance along this direction is preserved.
The partition function will therefor have an infinite factor
coming from this direction.

The reasonable thing to do is therefore to compactify that direction
as well.
We end up with M-theory on $\MT{9}$ with $Spin(2) = U(1)$ twists 
of the transverse $\MR{2}$.

M-theory on $\MT{9}$ is a two-dimensional theory and behaves drastically
different from the higher dimensional toroidal compactifications.
For one, the lengths of the sides of $\MT{9}$, as well as the fluxes along
it, are no longer moduli but are fluctuating fields.
The usual questions that we ask in 2+1D and higher, i.e. what
is the S-matrix of massless and massive particles, no longer make sense.
For one, there are probably no massive stable particles below
3+1D and moreover, in 1+1D the theory is conformally invariant.
I do not wish to imply that there are no interesting questions in 1+1D but
we probably have to work harder to find them.

Going back to our setting, we recall that
the usual argument for why there are no moduli is that 
small fluctuations in 1+1D have correlation functions which behave
like $\log r$ and do not fall off with distance.
However, our setting is not really two-dimensional!
As we have seen above, for generic irrational twists, the points 
far out at infinity can be surrounded with an arbitrarily large
11-dimensional ball.
Thus, it is plausible that the lengths, angles and fluxes of the
$\MT{9}$ are stable. They are determined by the boundary conditions at
infinity.
Let us therefore assume that such a partition function, $\ZM_9$, exists.

One can show that the partition function $\ZM_7$ of M-theory
on $\MT{7}$ with $Spin(4)$ twists is encoded in $\ZM_9$.
We can decompactify two sides of $\MT{9}$ to get $\MT{7}$ but
at first sight it seems that the twists are restricted
to be in a $Spin(2)\subset Spin(4)$ subgroup.
How can we turn on the other parts of the twists?
The solution is the same one that was used in \cite{CGKM}.
We first note the following (see \cite{CGKM} for more details).
Take M-theory on a large $\MS{1}$ and write
down the classical solution for a Kaluza-Klein monopole
which is homogeneous in 6+1D and has 3 transverse directions other
than the circle.
At the origin of the solution the translations along $\MS{1}$
get mapped to a $U(1)$ subgroup of one of the $SU(2)$ factors
of the $SO(4)$ rotation of the tangent space.
More precisely, there is an isometry of the space which at infinity
acts as translations of $\MS{1}$
and at the origin acts on the tangent space
as a twist in one $SU(2)$ factor.
There is also another $SO(3)$ isometry  group
that rotates the transverse $\MR{3}$ at infinity and acts as 
the second  $SU(2)$ factor (or rather the diagonal subgroup
of the two $SU(2)$'s) at the origin.

Thus if we compactify on $\MT{7}\times\MS{1}\times\MS{1}$
with $U(1)$ twists and take
the radii $R_7,R_8$ of the two extra $\MS{1}$'s to be much larger
than the size of $\MT{7}$ and much larger than the Planck length $M_p^{-1}$,
and if we also take $R_8\gg R_7$ then we can read off $\ZM_7$ with
generic $Spin(4)$ twists as follows.
We separate from $\ZM_9$ the part that has a Kaluza-Klein (KK) monopole
with respect to the first $\MS{1}$ factor (of radius $R_7$). The 
KK-monopole is required to
be filling $\MT{7}$. Up to a minor subtlety that will 
be discussed later on, we isolate this part from $\ZM_9$
by Fourier decomposing $\ZM_9$
with respect to the phase $\chi$ which couples to the KK monopole
(i.e. the 2+1D dual of the field $g_{\u 7}$).
Note that the tension of the KK monopole is $R_7^2 M_p^9$
which is much larger than all the other branes in this limit.
Thus the fluctuations of the KK-monopole itself are negligible in
this limit.
The missing twists (in the missing $SU(2)$ factor)
are simply rotations in the $\MS{1}$ as one
goes around cycles of $\MT{7}$.
Moreover, as in \cite{CGKM},
we can obtain the $(2,0)$ and little-string theories
of M5 and NS5-branes at $A_{q-1}$ singularities \cite{BluInt,IntNEW}
by taking $q$ KK-monopoles instead of one.
In our case, these come with a phase $e^{i q\chi}$ 
and can be read off-from $\ZM_9$ by a Fourier decomposition.
One can obtain the quiver theories of \cite{DouMoo} by compactifying
these 5+1D theories on a small $\MT{2}$.

\subsection{From $E_9$ to $E_{10}$}
At this point the function $\ZM_9$ depends on the parameter 
space which heuristically looks like the moduli of $\MT{9}$
and the periodic twists. In higher dimensions,
we saw that the twists are related to the charges of M-theory
on $\MT{d}$. Let us digress to discuss what happens
to the moduli space of M-theory on $\MT{d}$ when we add circles (to
the moduli space)
corresponding to periodic variables which are labeled by the charges. 
The moduli space is of the form
$E_{d(d)}(\BZ)\backslash E_{d(d)}(\BR)/\Kom$ where $\Kom$ is a maximal
compact subgroup of $E_{d(d)}(\BR)$ \cite{CJ,HulTow}.
Mathematically, adding the periodic variables can be achieved by
going from  $d$ to $d+1$, i.e. by compactifying on an extra
$\MS{1}$ and identifying the periodic variables with Wilson lines
along the extra $\MS{1}$.
There is also one extra variable -- the size of $\MS{1}$ which we 
will set to infinity.

In our case we would like to say that $\ZM_9$ depends on
$E_9(\BR)/\Kom_9$ and the extra periodic variables corresponding to the
twist. Because the dimension is so low, it is not clear
what are the possible twists. Our rule that there is a twist  for 
every charge in M-theory on $\MT{d}$ does not apply here because
in 2+1D and lower there are no BPS charges for local symmetries.
However, according to the above discussion there is a natural
conjecture.
The space $\MTW_9$, that is heuristically $E_9$ with periodic variables
corresponding to ``charges'', should be identified with,
\be\label{etenm}
E_{10}(\BZ)\backslash E_{10}(\BR)/\Kom,
\ee
where $\Kom$ is the maximal subgroup of $E_{10}$!
The mysterious $E_{10}$ group has appeared in conjectures about M-theory
in \cite{NicEX,HulTow,MartEX,EGKR,HarMoo} and more.
This infinite dimensional group appears in discussions about
M-theory on $\MT{10}$.
However, when one talks about M-theory on $\MT{10}$, the notion
of ``moduli-space'' is not well-defined and it is not
completely clear to me what exactly $E_{10}$ means in that context.

In our setting we appear to be avoiding this problem because the 
M-theory background is actually 11-dimensional (for generic irrational
twists).
Because points which are far out in space-time can be surrounded by
arbitrarily large balls, the correlation functions for small fluctuations
around VEVs are not 2-dimensional and do fall off.

So, we end this subsection with the conjecture that there exists a certain
function $\ZM$ on (\ref{etenm}) that encodes all the partition functions
of gauge theories, $(2,0)$ and little-string theories as well
as some information about S-matrices in M-theory on $\MT{d}$ for $d\le 8$.

Later on, we will conjecture that this function satisfies an
(infinite-dimensional) Laplace equation $\nabla \ZM = 0$
similarly to \cite{GGV,PiolH,GreSet,Piol}.

\subsection{An example}
We will end this section with an example of the partition
function of a field theory on a space with a geometrical twist.
We will study the partition function of a free scalar field
on an $\MR{2}$ bundle over $\MT{2}$ with two twists,
$$
g_1,g_2\in U(1),\qquad g_i \equiv e^{i\a_i}.
$$
$U(1)$ here is the rotations of the fiber $\MR{2}$.
Because $\MR{2}$ is non-compact we need to regularize.
One way to try to regularize might be to replace $\MR{2}$ with
a large $\MS{2}$ of radius $R$
and then divide by two because of the contribution of the two poles,
however I was not able to proceed with this approach.
Instead, we will  propose another regularization.
Let us set $\a_1=0$, for simplicity, and let us assume that
$\a_2$ is an irrational number.
The partition function is then,
$$
Z = \tr{e^{-R_2 H} \Omega(\a_2)},
$$
where $H$ is the Hamiltonian and $\Omega$ is the operator of rotation
of the fiber $\MR{2}$.
Now let us expand in a Fock space of free particle states with
definite momentum along $\MR{2}$.
We get,
$$
\sum_{r=0}^\infty {1\over {r!}}\sum_{p_1,\dots p_r}e^{-T\sum |p_j|}
\langle p_1,p_2,\dots,p_r |
\Omega p_1,\Omega p_2,\dots,\Omega p_r\rangle
$$
Because $\a_2$ is irrational, the only way that the matrix element
will not vanish is that all $p_i = 0$.
(For rational $\a_2 = {p\over q}$, it is possible to have a nonvanishing
matrix element if $r$ is a multiple of $q$.)
For $r=0$ the scalars contribute $1$. 
Using,
$$
\int d^2 k \langle k | \Omega(\a) k\rangle
   = \det (\Id - \Omega)^{-1} = {1\over {4\sin^2{\pi\a}}},
$$
we obtain the single particle ($r=1$) contribution for scalars,
$$
\det(\Id-\Omega)^{-1}\sum_{k} e^{-|k|\tau_2 -ik\tau_1},
$$
where $\tau = \tau_1 + i\tau_2$ is the complex structure of the $\MT{2}$.
Now let us analyze the 2-particle states.
We have,
\bear
Z_1 &=&
{1\over 2}\sum_{k_1,k_2}
e^{-(|k|_1 + |k|_2) \tau_2 + i(k_1 + k_2)\tau_1}\nn\\
&& \int d^2 p_1\, d^2 p_2\,
(\delta^{(2)}(p_1 - \Omega p_1)\delta^{(2)}(p_2 - \Omega p_2)
\nn\\ && + \delta_{k_1,k_2}
\delta^{(2)}(p_1 - \Omega p_2)\delta^{(2)}(p_2 - \Omega p_1))
\nn\\ &=&
{1\over 2}\left(\det(\Id-\Omega)^{-1}\sum_k
e^{-|k|\tau_2 -ik\tau_1}\right)^2
+{1\over 2}\det(\Id-\Omega^2)^{-1}\sum_{k}
e^{-2|k|\tau_2 -2ik\tau_1}.
\label{ettpwa}
\eear
Let us denote,
$$
F^{(0)}_n(\a,\tau)\equiv \det(\Id - \Omega(\a)^n)^{-1}
    {\sum_k}' e^{-n|k|\tau_2 - i n k\tau_1}.
$$
Note that we have excluded the $k = 0$ term from the sum.
This term is the contribution of the zero
mode of the scalar field (since the momenta in the $\MR{2}$
direction are also zero) and we have to exclude it since we have fixed
the VEV.
Now, for the $r$-particle sector, we should choose a permutation in
$S_r$ with $\sigma_l$ cycles of length $l$,
$$
\sum_{l} l\sigma_l = r.
$$
The contribution of this permutation is,
$$
{1\over {r!}}\prod_l F_l(\a\tau)^{\sigma_l}.
$$
The number of such permutations is,
$$
{{r!}\over {\prod \sigma_l!
\prod (l!)^{\sigma_l}}}\prod ((l-1)!)^{\sigma_l}
 = {{r!}\over {\prod l^{\sigma_l}\sigma_l!}}
$$
Thus, the contribution of the scalar fields to the regularized partition
function is,
$$
e^{\sum_{l=0}^\infty {{F^{(0)}_l(\a,\tau)}\over l}}.
$$
Note that $F_n(\a)$ has a pole for $\a= {m\over n}$.
For generic irrational $\a$ the sum $\sum {{F_n}\over n}$
probably converges. Even though transcendental numbers
can approximate  rational numbers faster than any power,
that is, for any $c$ there exist arbitrarily large $N$'s such that
$|e^{2\pi i N\a} - 1| < {1\over {N^c}}$, the exponential term
in $F_l$ can suppress the large $\det^{-1}(\Id-\Omega^N)$.

\subsubsection*{Twists along both cycles}
Now we change to a basis $|k,m,q\rangle$ where $m$ is the angular
momentum along the fiber $\MR{2}$ and $q\ge 0$ is the radial momentum.
The energy of the state is,
$$
{\cal E}(k,m,q) = \sqrt{q^2 + \left({{k + m\a_1}\over {R_1}}\right)^2}.
$$
The states are normalized as,
$$
\langle k',m',q' | k,m,q\rangle = 
{1\over q}\delta_{m'm}\delta_{k'k}\delta_{q-q'},
$$
so that the identity operator is,
$$
\Id = \sum_m \sum_k\int q dq|k,m,q\rangle\langle k,m,q|.
$$
The operator $\Omega e^{-R_2 H}$ is diagonal in
this basis and the partition function becomes,
$$
\log Z = R_2^2\sum_m\sum_k\int q dq 
 \log\left(1 
   -e^{-\sqrt{q^2 + (k + m\a_1)^2\tau_2^2} - i k\tau_1-im\a_2}\right).
$$
We will not discuss modular invariance here, but in general
we can work only with regularizations that give a modular invariant
result.


\section{Lie algebras and Lie groups}
The moduli space of M-theory on $\MT{d}$ (with $d\le 8$) is
given by the coset \cite{HulTow},
$$
E_{d(d)}(\BZ)\backslash E_{d(d)}(\BR) / \Kom,
$$
where $\Kom$ is the maximal compact subgroup.
We will now review some mathematical facts about these cosets
and their relation to the moduli spaces of M-theory on $\MT{d}$.

We will start by reviewing the construction of Lie algebras and Lie
groups (see for instance \cite{Kac}).
We will also explain the distinction between the compact
and noncompact real versions of the groups. 
We will use $E_8$ and $E_{8(8)}$ for demonstration.

\subsection{Roots and Chevalley generators}
The  Lie algebras of both $E_8$ and $E_{8(8)}$ contain 248 generators
$\tau^a$ ($a=1\dots 248$) and if we allow combinations with complex
coefficients the two \emph{complexified} Lie algebras are the same.
This means that there is a $248\times 248$ matrix with complex 
coefficients which transforms the commutation relations of $E_8$ to
those of $E_{8(8)}$ but there is no such map with real coefficients.

Let us describe the complexified Lie algebra first.
It has a root lattice $\Delta\subset \calh^* =\MR{8}$
given by ($v_1\dots v_8$
are orthogonal unit vectors in $\MR{8}$):
\be\label{eroots}
\Delta = \{\pm v_i\pm v_j,
\half (\pm v_1\pm v_2\pm\cdots \pm v_8)\qquad 
({\rm even\ number\ of\ minuses})\},
\ee
and the 240 simple roots are given by
\bear
\Pi &=& \{
\a_1 = v_2 - v_3,   \a_2 = v_3 - v_4, \a_3 = v_4 - v_5,    \nn\\
&&\a_4 = v_5 - v_6,   \a_5 = v_6 - v_7,   \a_6 = v_7 - v_8, \nn\\
&&\a_7 = \half (v_1 - v_2 - \cdots - v_7 + v_8), \a_8 = v_7 + v_8\}.
\label{simroo}
\eear
For future reference we also write down the highest weight,
\be\label{rootth}
\theta = v_1 + v_2 = 2\a_1 + 3\a_2 + 4\a_3 + 5\a_4 + 6\a_5 
                   + 4\a_6 + 2\a_7 + 3\a_8.
\ee
The Lie algebra is generated as follows.
We take the dual $\calh = \MR{8}$ and choose basis vectors
$$
\a_1^\vee,\dots,\a_8^\vee\in\calh,
$$
such that
$$
a_{ij} = \lrp{\a_i^\vee}{\a_j},
$$
is the Cartan matrix.
We now define 8 pairs of Chevalley generators 
$$
e_1, f_1, \dots,e_8, f_8
$$
with the commutation relations, 
\bear
\com{e_i}{f_j} &=& \delta_{ij} \a_i^\vee\in\calh,\qquad i,j=1\dots 8,\nn\\
\com{h}{e_i} &=& \lrp{\a_i}{h}
                e_i,\qquad h\in\calh,\qquad i=1\dots 8,\nn\\
\com{h}{f_i} &=& -\lrp{\a_i}{h}
                 f_i,\qquad h\in\calh,\qquad i=1\dots 8.\nn\\
&&\label{chevcom}
\eear
We supplement the relations with the conditions that for $i\neq j$,
$$
({\mbox{ad}}\, e_i)^{1-a_{ij}} e_j = 0,\qquad
({\mbox{ad}}\, f_i)^{1-a_{ij}} f_j = 0,
$$
where ``${\mbox{ad}}\, x$'' is defined by,
$$
({\mbox{ad}}\, x)^k y \equiv
\underbrace{\lbrack x, \lbrack x,\cdots\lbrack x,}_{k\ {\rm times}} y
\rbrack\cdots\rbrack\rbrack.
$$

We can also separate the generators $\tau^a$ into positive ($\tau^{+\a}$),
negative ($\tau^{-\a}$) and those belonging to the Cartan subalgebra
($\tau^i$). For $E_8$ we get  $8+120+120$ generators as follows,
$$
\tau^i = \a_i^\vee,\qquad i=1\dots 8,\qquad
\tau^{+\a},\qquad \a=1\dots 120,\qquad
\tau^{-\a},\qquad \a=1\dots 120.
$$
In our notations $\a\in\Delta_{+}$ will always be a positive root
(a combination of $\a_1\dots\a_8$ with positive integers) and 
$-\a\in\Delta_{-}$ will be a negative root.

The commutation relations are derived from
$$
\com{\tau^i}{\tau^{+\a}} = \lrp{\a_i^\vee}{\a} \tau^{+\a},\qquad
\com{\tau^i}{\tau^{-\a}} = -\lrp{\a_i^\vee}{\a} \tau^{-\a},\qquad
$$
Also, for the simple roots,
$$
\tau^{\a_i} = e_i,\qquad
\tau^{-\a_i} = f_i,\qquad i=1\dots 8.
$$

\subsection{Reality compactness and maximally split form}
To describe the real groups $E_8$ and $E_{8(8)}$ we have to specify
what are the restrictions on the 248 coefficients $c_i, c_{\a}, c_{-\a}$
such that an expression of the form,
$$
\tau = \sum_{i=1}^8 c_i\tau^i + \sum_{\a\in\Delta_{+}} 
       (C_{\a}\tau^{+\a} + C_{-\a}\tau^{-\a}),
$$
will be a generator of the corresponding real Lie algebra.
We will then define the real group to be represented
by the matrices $e^{\tau}$ (in the adjoint representation).

To specify which coefficients correspond to a real generator of the
compact $E_8$ or of the non-compact $E_{8(8)}$, we have to define
an involution (a different one for $E_8$ and $E_{8(8)}$).
 Acting on the complexified Lie algebra $\calg$ it satisfies,
\be\label{chinvdef}
\omega\{\com{x}{y}\} = \com{\omega x}{\omega y},\qquad
\omega\{\lam x\} = \bar{\lam} \omega x,\qquad
x,y\in\calg,\qquad \lam\in\BC.
\ee
It is sufficient to define $\omega$ on the Chevalley 
generators $e_i, f_i, h_i$.
The real Lie algebra will then be defined by,
$$
\calg_R = \{x\in\calg | \omega x = x\}.
$$
For the compact $E_8$, we
define the compact Chevalley involution $\omega_c$ as
(p25 of \cite{Kac}),
$$
\omega_c(h_i) = -h_i,\qquad
\omega_c(e_i) = -f_i,\qquad
\omega_c(f_i) = -e_i,
$$
In particular, expressions like
$$
{\mbox{\rm exp}}
\left\{i \sum_1^8 c_i h_i\right\},\qquad c_1,\dots c_8\in\BR
$$
are in the compact group.

For the noncompact $E_{8(8)}$, there are several
ways to define $\omega_n$ depending
on whether the $h_i$ define compact or non-compact directions.
We choose to define the involution $\omega_n$ as
\be\label{chevn}
\omega_n(h_i) = h_i,\qquad
\omega_n(e_i) = e_i,\qquad
\omega_n(f_i) = f_i,
\ee
In particular, expressions like
\be\label{explike}
{\mbox{\rm exp}}
\left\{\sum_1^8 c_i h_i\right\},\qquad c_1,\dots c_8\in\BR,
\ee
are in the noncompact group.

\subsection{The Weyl group}
The Weyl subgroup, $\Weyl$,
 is generated by the reflections around the roots,
$$
\sigma_\a(x) = x - \lrp{x}{\a} \a,\qquad x\in\calh^*.
$$
Here, $\a$ is a root and $x$ is in the space generated by the roots.
Recall that
$\lrp{\a}{\b} = {{2(\b,\a)}\over {(\a,\a)}}$,
but we will always work with simply-laced groups for which
$(\a,\a) = 2$ for all the roots.
For finite simply-laced groups and any two roots $\a$ and $\b$
one can find $\sigma\in\Weyl$ with $\sigma(\a)=\b$.
The order of the Weyl group for $E_8$ is
$2^{14}\times 3^4\times 5^2\times 7$.

\subsection{$\Nil\Abl\Kom$ decomposition}
The maximally split real groups like $SL(n,\BR)$ and $E_{8(8)}$
have a decomposition into a nilpotent subgroup times an abelian subgroup
times a compact subgroup.
For example, an element $g\in SL(2,\BR)$ can be uniquely decomposed as,
$$
g = \Nil\Abl\Kom = 
\left(\begin{array}{rr} 1 & a \\ 0 & 1 \\  \end{array}\right)
\left(\begin{array}{rr} e^\lam & 0 \\ 0 & e^{-\lam} \\ \end{array}\right)
\left(\begin{array}{rr}
        \cos\tht & \sin\tht \\
       -\sin\tht & \cos\tht \\
\end{array}\right).
$$
The $SL(2,\BR)$-invariant metric can be written as,
$$
ds^2 = \tr{g^{-1} dg g^{-1} dg} = 
{1\over 2}e^{-4\lam} da^2 + 2 d\lam^2
  - 2 (d\tht + {1\over 2}e^{-2\lam}da)^2.
$$
An element in the coset $SL(2,\BR)/SO(2)$ can be uniquely represented as,
$\Nil\Abl\Kom /\Kom = \Nil\Abl$. It can be parameterized by $\lam$ and $a$.
The metric is,
$$
ds^2 = 
{1\over 2}e^{-4\lam} da^2 + 2 d\lam^2.
$$
An element in the coset $SL(3,\BR)/SO(3)$
can be decomposed as $\Nil\Abl$ with,

\be\label{nilabl}
\Nil = 
\left(\begin{array}{rrr}
1 & a & b \\
0 & 1 & c \\
0 & 0 & 1 \\
\end{array}\right),\qquad
\Abl =
\left(\begin{array}{rrr}
e^\lam & 0 & 0 \\
0 & e^\mu & 0 \\
0 & 0 & e^{-\lam-\mu} \\
\end{array}\right).
\ee
The $SL(3,\BR)$-invariant metric can be written as,
\bear
ds^2 &=&
2 d{\lambda}^2 
+ 2 d{\lambda} {d\mu} 
+ 2 d{\mu}^2 \nn\\
&& + {1\over 2} e^{2 \mu - 2 \lambda} da^2
+ {1\over 2} e^{-2 \lambda - 4 \mu} dc^2
+ {1\over 2} e^{-4 \lambda - 2 \mu} (db - c da)^2.
\label{metcoslth}
\eear
In general, for a real Lie-algebra defined by the involution $\omega_n$
above, we can decompose an element $g\in G$ as,
\be\label{gcalp}
g = e^{\sum_{\a\in\Delta_+} C_{\a}\tau^{+\a}} e^\lam
 e^{i \sum_{\a\in\Delta_+} B_{\a} (\tau^{+\a} - \omega_n(\tau^{+\a}))},
\ee
where $\lam\in\calh$ and $C_{\a}$ and $B_{\a}$ are real.
We can also normalize $\tau^{-\a}= \omega_n(\tau^{+\a})$.
The compact subgroup $K$ is generated by,
$$
{\mbox{\rm exp}}\left\{
i \sum_{\a\in\Delta_+} B_{\a} (\tau^{+\a} - \omega_n(\tau^{+\a}))
\right\}.
$$
To first order in $c_\a$ and any order in $\lam$, we can calculate the
$G$-invariant metric on $G/K$ to be,
\be\label{metna}
ds^2 = \| d\lam\|^2 
+ {1\over 2}\sum_{\a\in\Delta_+} e^{-2\lrp{\lam}{\a}} dC_{\a}^2
+\cdots
\ee

We will end this section with the formula for the $SL(n,R)$
invariant Laplacian on the coset $SL(n,\BR)/SO(n)$.
An element $g\in SL(n,\BR)$ can be written as $N A K$
with $K\in SO(n)$ and,
\be\label{naslthr}
N = \left(\begin{array}{ccccc}
1 & C_{12} & C_{13} & \cdots & C_{1,n-1} \\
0 & 1 & C_{22} &  \cdots & C_{2,n-1} \\
\vdots & \vdots & \ddots &  \vdots & \vdots \\
0 & 0 & 0 &  \cdots & 1 \\
\end{array}\right),\qquad
A = \left(\begin{array}{ccccc}
e^{-\mu_1} & 0 & 0 & \cdots & 0 \\
0 & e^{-\mu_2} & 0 & \cdots & 0 \\
\vdots & \vdots & \ddots & \vdots & \vdots \\
0 & 0 & 0 & \cdots & e^{-\mu_n} \\
\end{array}\right).
\ee

Here we denoted,
$$
\mu_n \equiv -\sum_{j=1}^{n-1}\mu_j.
$$
The metric can be written as,
$$
ds^2 = \tr{(A^{-1} dA)^2} 
+ {1\over 2}\tr{A^{-2} N^{-1} dN A^2 dN^T (N^{-1})^T}.
$$
The Laplacian becomes,
\bear
\nabla &=& \sum_{k=1}^{n-1} {{\partial^2}\over {\partial\mu_k^2}}
-{1\over n}\left(\sum_{k=1}^{n-1} {{\partial}\over {\partial\mu_k}}\right)^2
+\sum_{k=1}^{n-1}(n+1-2k){{\partial}\over {\partial\mu_k}}
\nn\\
&+&
{1\over 2}\sum_{1\le i<j\le n} e^{2\mu_j-2\mu_i}
     {{\partial^2}\over {\partial C_{ij}^2}}
+\sum_{1\le i<k<j\le n} e^{2\mu_j-\mu_k-\mu_i}
     \left(C_{ik} +\sum_{1\le l<i} C_{li} C_{lk}\right)
{{\partial^2}\over {\partial C_{kj} \partial C_{ij}}}.
\nn\\ &&
\eear


\section{The moduli space of M-theory on $\MT{8}$}
In this section we will briefly review the relation between
$E_{8(8)}/SO(16)$ and M-theory on $\MT{8}$. 
 A comprehensive review appears in \cite{OPRev,OPRevSh,OPNew}.

\subsection{The compactification radii}
Let us denote $\Grp\equiv E_{8(8)}(\BR)$, $\Kom\equiv SO(16)$ and
$\Gamma\equiv E_{8(8)}(\BZ)$. We will also denote
the maximal abelian subgroup by $\Abl\sim SO(1,1)^8$, as before.
The moduli space of M-theory on $\MT{8}$ is $\Gamma\backslash \Grp/\Kom$
\cite{MarSch}. Using the $\Grp=\Nil\Abl\Kom$ decomposition,
the moduli-space can be written as $\Gamma\backslash \Nil\Abl$.
A sublocus of this moduli-space is given by compactification on
tori which are a direct product,
$$
\MT{8} = \MS{1}\times\cdots\times\MS{1},
$$
of circles of radii $R_1,\dots,R_8$ and with all 3-form and 6-form fluxes
turned to zero as well as the duals of all vectors  turned to zero.
Let,
\be\label{radii}
R_i = e^{t_i},\qquad i=1\dots 8,
\ee
be the radii in 10+1D M-theory units.
Up to an element in the U-duality group $E_{8(8)}(\BZ)$, the corresponding
point in the moduli space can be defined as having $\Nil=\Id$.
The element in $\Abl$ can be taken to be,
$$
{\mbox{exp}}\{ s_1 h_1 + s_2 h_2 + \cdots + s_8 h_8\},
$$
where $h_i$ is the element in $\calh$ which corresponds to $\a_i^\vee$
in (\ref{simroo}) and the $s_i$ are related to $t_i \equiv \log R_i$ by,
\be\label{stot}
s_1 = t_1 - t_2,\,\,
s_2 = t_2 - t_3,\dots,
s_7 = t_7 - t_8,\,\,
s_8 = t_6 + t_7 + t_8.
\ee

\subsection{U-duality and the Weyl group}
The subgroup $\Gamma_T\subset \Gamma$ which preserves the condition
$\Nil = \Id$ is the subgroup of the U-duality group that preserves 
the structure of an orthogonal $\MT{8}$ with no fluxes.
It is generated by the permutation group $S_8$ acting on the 
8 radii $R_8$ and by T-duality which takes $R_1\dots R_8$ to
\be\label{tdl}
(R_6 R_7 R_8)^{1/3} R_1,\dots
(R_6 R_7 R_8)^{1/3} R_5,\,
(R_6 R_7 R_8)^{-2/3} R_6,\,
(R_6 R_7 R_8)^{-2/3} R_7,\,
(R_6 R_7 R_8)^{-2/3} R_8.
\ee
The claim is (see \cite{EGKR}) that this subgroup $\Gamma_T$ is
given by the Weyl subgroup $\Weyl\subset \Grp$ and also 
that actually $\Weyl\subset \Kom$.
By definition $\Weyl$ is constructed as follows.
Let $M\subset \Grp$ be the subgroup 
of all $w\in \Grp$ which keep $\Abl$ invariant ($w \Abl w^{-1} = \Abl$).
$\Abl$ is a normal subgroup of $M$ and $\Weyl = M/\Abl$.

The vector,
$$
u = (\log R_1,\dots,\log R_8),
$$
is a linear representation of the Weyl group.
It can be checked that
if an element $w\in \Weyl$ acts on the simple root $\a_j$ as
$\Lambda_{jk}\a_k$
then it acts on $u$ as $B \Lambda B^{-1}$ where $B$ is the matrix,
\be\label{bmat}
B = \left(\begin{array}{rrrrrrrr}
    -1 &  0 &  0 &  0 &  0 &  0 &  0 &  0\\
     1 & -1 &  0 &  0 &  0 &  0 &  0 &  0\\
     0 &  1 & -1 &  0 &  0 &  0 &  0 &  0\\
     0 &  0 &  1 & -1 &  0 &  0 &  0 &  0\\
     0 &  0 &  0 &  1 & -1 &  0 &  0 &  0\\
     0 &  0 &  0 &  0 &  1 & -1 &  0 & -1\\
     0 &  0 &  0 &  0 &  0 &  1 & -1 & -1\\
     0 &  0 &  0 &  0 &  0 &  0 &  1 & -1\\
\end{array}\right).
\ee

\subsection{Harmonic functions and masses of BPS states}
M-theory on $\MT{d}$ for $d\le 7$ has $U(1)$ charges
in $\MR{10-d,1}$ which are in a representation of the U-duality
group $E_{d(d)}(\BZ)$. There are also BPS states which are charged
under these $U(1)$'s. The mass formula that relates the mass of the BPS
state to its charge and the point in the moduli-space was derived
in \cite{WitVAR}.
M-theory on $\MT{8}$ is three-dimensional and does not possess $U(1)$
charges in the usual sense. On the other hand, we  can get and instanton
by taking a 3+1D BPS state of M-theory
on $\MT{7}$ and letting it have a Euclidean world-line
inside $\MT{8}$. Any given instanton makes sense when going to a particular
limit in the moduli space $E_{8(8)}(\BR)/SO(16)$ for which the action
of the instanton becomes infinite. In that limit one can make
an instanton expansion for the contribution of the instanton to certain
low-energy quantities.
In \cite{GGV},
the instanton expansion for type-IIB was studied.
It was proposed that certain terms in the effective 9+1D low-energy
description have a coefficient which is a harmonic (up to a constant)
function on the moduli-space parameterized by the complex dilaton
$\tau$. In the limit $\Im\tau\rightarrow \infty$, the instanton expansion
was recovered. This conjecture was generalized in
\cite{KPI},\cite{APT},\cite{KPII}-\cite{Piol}
to compactifications up to $\MT{4}$.
Furthermore, in \cite{GreSet} it was proved that the coefficients
of the 16-fermion terms must be harmonic (up to a constant)
in order for type-IIB in 9+1D to be supersymmetric.

Although the details of the corrections to
the low-energy effective action of M-theory on $\MT{8}$ have never,
as far as I know, been worked out in detail, it is natural to
believe that they too are given by a harmonic function on the moduli-space.
Because this is going to be important later on,
let us explain the details of how this works.
We assume that $E_{8(8)}/SO(16)$ is parameterized as $\Nil\Abl$ (as above)
and that $\Abl$ is parameterized by $\lam\in\calh^*$ and $\Nil$ is
parameterized by the $C_{\a}$'s.
We will use the approximate expression (\ref{metna}) for the metric.
Pick a particular positive root $\a$.
For regions in the moduli-space for  which
$-\lrp{\lam}{\a} \gg 1$ the function,
\be\label{incon}
\Psi(\{C_{\a}\},\lam) =
e^{-2\pi |n|e^{-\lrp{\lam}{\a}} + 2\pi i n C_\a},
\ee
is an approximate harmonic function.
The U-duality group $E_{8(8)}$ forces the identification
$C_{\a}\sim C_{\a}+1$ (when all other $C_{\b}$'s are zero) and thus
$n$ is an integer.
The factor (\ref{incon})
can be interpreted as the contribution of an instanton
made by a brane which is wrapped $n$ times.
Let us go over the simple roots and see what kind of instantons
we get.
Using (\ref{stot}) and (\ref{simroo}) we find that the 120 positive roots
of $E_{8(8)}$ correspond to the following list.
There are 28 instantons made by KK particles
with Euclidean world-lines. Their actions are,
$$
e^{t_i - t_j} = R_i R_j^{-1},\qquad 1\le i<j\le 8.
$$
There are 56 M2-branes with actions,
$$
e^{t_i + t_j + t_k} = R_i R_j R_k,\qquad 1\le i<j<k\le 8.
$$
There are 28 M5-branes with actions,
$$
e^{t_i + t_j + t_k + t_l + t_m + t_n} = R_i R_j R_k R_l R_m R_n,\qquad
1\le i<j<k<l<m<n\le 8.
$$
Finally, there are 8 KK-monopoles with actions,
$$
e^{t_i + \sum_{j=1}^8 t_j} = R_i\prod_{j=1}^8 R_j,\qquad j=1\dots 8.
$$
The highest weight $\theta = v_1 + v_2$ corresponds to a KK monopole
with respect to the first circle, with an action,
$$
R_1^2 R_2 R_3 R_4 R_5 R_6 R_7 R_8.
$$
For future reference, we will explicitly write the roots 
corresponding to a few states.
\bear
e^{t_1 + t_2 + t_3 + t_4 + t_5 + t_6} &\rightarrow&
 \b'_0\equiv
\a_1 +2\a_2 +3\a_3 +4\a_4 +5\a_5 + 4\a_6 + 2\a_7 + 2\a_8 = v_1 - v_8,
\label{defbzerp}\\
e^{t_7 + \sum_1^8 t_j} &\rightarrow&
 \a_1 +2\a_2 +3\a_3 +4\a_4 +5\a_5 +3\a_6 +2\a_7 +3\a_8 = v_1 + v_8,\nn\\
e^{t_8 + \sum_1^8 t_j} &\rightarrow&
 \b'_1\equiv \a_1 +2\a_2 +3\a_3 +4\a_4 +5\a_5 +3\a_6 +\a_7 +3\a_8 
  \nn\\ &&= {1\over 2}(v_1 + v_2 + v_3 + v_4 + v_5 + v_6 + v_7 + v_8).
\label{defbonp}
\eear
The first is an M5-brane and the last two are KK-monopoles.
We will also need the roots corresponding to KK states along
directions $1\dots 6$,
\bear
e^{t_k - t_7} &\rightarrow& \sum_{j=k}^6\a_j = v_{k+1} - v_8,
\qquad k=1\dots 6,\nn\\
e^{t_k - t_8} &\rightarrow& \gamma_k\equiv \sum_{j=k}^7\a_j =
v_{k+1} + {1\over 2}
    (v_1 - v_2 - v_3 - v_4 - v_5 - v_6 + v_7 + v_8),
\qquad k=1\dots 6.\nn\\
&&\label{morkkinp}
\eear

\subsection{Chern-Simons terms
 and the structure of $\Gamma\backslash\Nil\Abl$}\label{subsec:phxcst}
One of the effects of dividing by $\Gamma=E_8(\BZ)$ on the left
is to make the variables $C_{\a}$ periodic.
In this subsection we will discuss the topology of the resulting
manifold. Understanding exactly how the variables are periodic
will be important later on and will play a crucial role in the
discussion of electric and magnetic fluxes.
In the discussion below we will restrict ourselves to simple cases.
For a more comprehensive discussion on these issues see
\cite{OPR}.

We start with the example of $SL(3,\BR)$ from the previous
section.
In conjunction with the
decomposition (\ref{nilabl}), let us take the following 3 elements
of $SL(3,\BZ)$,
\be
U_1 =
\left(\begin{array}{rrr}
1 & 1 & 0 \\
0 & 1 & 0 \\
0 & 0 & 1 \\
\end{array}\right),\qquad
U_2 =
\left(\begin{array}{rrr}
1 & 0 & 0 \\
0 & 1 & 1 \\
0 & 0 & 1 \\
\end{array}\right),\qquad
U_3 = U_1 U_2 U_1^{-1} U_2^{-1} = 
\left(\begin{array}{rrr}
1 & 0 & 1 \\
0 & 1 & 0 \\
0 & 0 & 1 \\
\end{array}\right).
\ee
They act on $a,b,c$ of~(\ref{naslthr}) as,
\bear
U_1 &:&\,\, a\rightarrow a+1,\,\, b\rightarrow b+c,\,\,
                 c\rightarrow c,\nn\\
U_2 &:&\,\, a\rightarrow a,\,\, b\rightarrow b,\,\,
                 c\rightarrow c+1,\nn\\
U_3 &:&\,\, a\rightarrow a,\,\, b\rightarrow b+1,\,\,
                 c\rightarrow c,\nn
\eear

We see that modding out $\Nil$ by the subgroup generated by $U_1,U_2$
gives a manifold which is not simply $\MT{3}$ but can be described
as an $\MS{1}$ bundle over $\MT{2}$ with a first Chern class equal to 1.
The base $\MT{2}$ can be parameterized by $(a,c)$ and the fiber $\MS{1}$
can be parameterized by $b$.

In the general case, let $\a,\b$ be two positive roots. Let 
$\gamma=\a+\b$. Assume that $\gamma\in\Delta_{+}$.
Then there exists an element $U_\a$ in the U-duality group that acts as,
\be\label{uacc}
U_{\a}:\,\,C_{\a}\rightarrow C_{\a} + 1,\,\,
C_{\b}\rightarrow C_{\b},\,\,
C_{\gamma}\rightarrow C_{\gamma} + {1\over 2}C_{\b},\,\,\cdots
\ee
Similarly, there exists an element $U_{\b}$ that acts as,
$$
U_{\b}:\,\,
C_{\a}\rightarrow C_{\a},\,\,
C_{\b}\rightarrow C_{\b} + 1,\,\,
C_{\gamma}\rightarrow C_{\gamma} - {1\over 2}C_{\a},\,\,\cdots
$$
(The factors of $\half$ appear because, as defined, the $C_\a$'s
appear in the exponent of~(\ref{gcalp}).)
The pair of coordinates $(C_{\a},C_{\b})$ with the identification
$C_{\a}\sim C_{\a}+1$ and $C_{\b}\sim C_{\b}+1$, parameterizes
a torus $\MT{2}$.
If we define,
\be\label{cgamh}
C'_{\gamma} = C_{\gamma} + {1\over 2}C_{\a} C_{\b},
\ee
then the phase $e^{i C^{(\pm)}_{\gamma}}$ is a section of a line
bundle over the $\MT{2}$ with $c_1 = 1$.
Note that we could just as well have chosen
the $(-)$ sign in the second term in~(\ref{cgamh}).
The sign is determined
by the normalization of the generator $\tau^{+\gamma}$
according to whether $\tau^{+\gamma} = \com{\tau^{+\a}}{\tau^{+\b}}$
or $\tau^{+\gamma} = \com{\tau^{+\b}}{\tau^{+\a}}$.

A special case of~(\ref{cgamh})
is when $C'_{\gamma}$ is the phase associated
with an M5-brane instanton wrapped on directions $1\dots 6$.
$$
e^{t_1 +t_2 +t_3 +t_4 +t_5 +t_6} \longrightarrow
\b'_0\equiv
\a_1 +2\a_2 +3\a_3 +4\a_4 +5\a_5 + 4\a_6 + 2\a_7 + 2\a_8.
$$
Let us take $\a$ and $\b$ to be the roots corresponding
to wrapped M2-branes as follows,
\bear
e^{t_4 +t_5 +t_6} &\longrightarrow&
  \a \equiv \a_4 + 2\a_5 + 2\a_6 + \a_7 + \a_8,\nn\\
e^{t_1 +t_2 +t_3} &\longrightarrow&
  \b \equiv \a_1 +2\a_2 +3\a_3 +3\a_4 +3\a_5 +2\a_6 +\a_7 +\a_8.\nn
\eear
In this case the statement above is a special case of the observation
made in \cite{WitFP} (and see also \cite{GanZER}) that the phase
of an M5-brane instanton is a section of a nontrivial circle-bundle
over the moduli-space of 3-form fluxes.

\subsection{The phase dual to the KK-monopole}\label{subsec:kkrev}
In the previous sections we argued that various partition functions
can be obtained from a function on the moduli space of $\MT{7}$
and the twists by Fourier transforming with respect to the phases.
Specifically, we needed to hook the contribution of a KK-monopole
with respect to the $8^{th}$ direction and filling directions
$1\dots 7$ together with Dehn twists in the $8^{th}$ direction
as one goes along one of the directions $1\dots 6$.
Let us denote by $\a'$ the root associated with the
phase corresponding to the KK-monopole,
$$
e^{t_8 + \sum_1^8 t_j} \rightarrow
 \a'\equiv \a_1 +2\a_2 +3\a_3 +4\a_4 +5\a_5 +3\a_6 +\a_7 +3\a_8,
$$
and by $\b'$ the phase corresponding to one of the Dehn twists, say,
$$
e^{t_1 - t_8}\longrightarrow
\b'\equiv \a_1+\a_2+\a_3+\a_4+\a_5+\a_6+\a_7.
$$
Now we calculate $\gamma'=\a'+\b'$ to be the phase corresponding
to another KK-monopole, this time with respect to the $1^{st}$ direction
and filling $2\dots 8$,
$$
t_1 + \sum_1^8 t_j =
 \gamma' = 2\a_1 +3\a_2 +4\a_3 +5\a_4 +6\a_5 +4\a_6 +2\a_7 +3\a_8.
$$
To find the instanton term we are looking for, we need to define,
$$
\ZM'(\cdots,\phi,\psi,\chi,\cdots)\equiv
\ZM(\cdots, C_{\a'} = \psi,\, C_{\b'} = \phi,\,
   C_{\gamma'} = \chi + {1\over 2} \psi\phi,\cdots).
$$
We then have to Fourier transform to get,
$$
\widetilde{\ZM}(\cdots,\phi,\cdots)\equiv \int d\psi\, d\chi\, e^{i\psi} 
\ZM'(\cdots,\phi,\psi,\chi,\cdots).
$$
This function will contain the M5-brane instanton term with the correct
dependence on the Dehn twist $\phi$.

\subsection{The phase dual to the M5-brane}\label{subsec:mfrev}
A similar situation also occurs for the phase of the M5-brane.
Let us take the phase $C_{\b'_0}$ (defined in (\ref{defbzero})
that is dual to the M5-brane charge and let us take $C_{\a_k}$ to correspond
to the Dehn twist of the $8^{th}$ direction along the $k^{th}$
(associated with the factor $e^{t_k - t_8}$).
Then $\b_k\equiv \b'_0 -\a_k$ is a positive root and $C_{\b_k}$
is dual to the charge of an M5-brane wrapped on the $8^{th}$ 
direction and on directions
$1\dots 6$ with the $k^{th}$ circle excluded.
As before, $e^{i C_{\b'_0}}$ is in a nontrivial line bundle over
the $(C_{\a_k},C_{\b_k})$ torus.
However, this time we cannot resolve the problem by integrating
over $C_{\b'_0}$ since we need to insert the factor $e^{i N C_{\b'_0}}$
which is not single-valued.

The resolution is to set $C_{\b_k} = 0$ instead.
Now, $C_{\b'_0}$ is single-valued (see (\ref{uacc}) with the definitions
$\gamma\equiv \b'_0$, $\a\equiv\a_k$ and $\b\equiv\b_k$).

This prescription also has a geometrical reason.
In the Taub-NUT metric corresponding to the KK-monopole,
the $8^{th}$ circle shrinks to zero at the origin of the transverse
space. The $N$ M5-branes are localized at that origin.
An M5-brane that would wrap the $8^{th}$ direction would also shrink
when it sits at the origin. This means that, in the presence of the
KK-monopole, the charge dual to $C_{\b_k}$ is not well-defined.
Thus, the only way out is to set $C_{\b_k} = 0$ so that there will
be no ambiguity due to phases of the form $e^{i C_{\b_k}}$.\footnote{
I have benefited from a discussion with M. Krogh on a similar 
setting in a different context.}

\subsection{Electric and Magnetic fluxes}\label{subsec:emflx}
As an application of (\ref{cgamh}) we will show how
to extract from $\ZM$
the sectors of gauge theories with electric and  magnetic
fluxes.
The discussion below is related to the discussion in
\cite{WitFP,AhaWit,WitFLX}. 

The ``partition-function'' on $\MT{4}$ (with $Spin(4)$ twists)
is not really a single function but depends on the gauge-bundle, in other
words, the electric and magnetic fluxes.
For $U(N)$ SYM the fluxes are characterized by an element of 
$H^2(\MT{4},\BZ)$ whereas for $SU(N)$ SYM they are characterized
by $H^2(\MT{4},\BZ_N)$. We will be working with the $U(N)$ theory.
To extract from $\ZM$ the sectors  with fluxes we just need to 
recall that inside a D3-brane $p$ units of electric flux have
the same quantum numbers as $p$ fundamental strings in the same direction
and $p$ units of magnetic flux have the same quantum numbers
as $p$ D1-branes in these directions (see \cite{SeiVBR}
and \cite{DouBWB}).
We now have two distinct ways to isolate the different sectors 
with fluxes from $\ZM$.
To isolate a partition function of a sector with $p$ units
of an electric flux
along the $1^{st}$ direction and the $4^{th}$ taken as time,
we need the phase,
$$
\sigma_{14} \equiv\int_{14} B^{(NSNS)},
$$
that couples to an instantonic
string wrapped on the $1^{st}$ and $6^{th}$ directions.
This phase is one of the  $C_{\a}$'s.
We then have to Fourier transform and find the coefficient
of $e^{i p\sigma_{14}}$.
Alternatively, we can treat another direction, say the $2^{nd}$ as time,
and isolate the phase corresponding to a Euclidean D1-brane
in directions $1,4$,
$$
\widetilde{\sigma_{14}} \equiv\int_{14} B^{(RR)}.
$$
Recall that in the $SU(N)$ case the two descriptions are related
by a discrete Fourier transform \cite{tHfFLX}.
In our case this follows from the two statements:
\begin{itemize}
\item
The partition function
$Z_N(\cdots,\sigma_{14},\widetilde{\sigma_{14}},\cdots)$
is a section of a line bundle with $c_1 = N$ over the torus that
is parameterized by
$(\sigma_{14},\widetilde{\sigma_{14}})$.
\item
$Z_N$ is a holomorphic in $\sigma_{14} + i\widetilde{\sigma_{14}}$.
\end{itemize}

Similar statements \cite{WitFP}
hold for the $(2,0)$ theory partition function
on $\MT{6}$ considered as a function of the phases,
$$
\int_{ijk} C_3,
$$
of the 3-form flux of M-theory in the M5-brane realization of the 
$(2,0)$ theory.
The partition function depends only on the self-dual combination
of the 3-form fluxes.
In all these cases, the fluxes correspond to $C_{\a}$ and 
$C_{\b}$ while the phase corresponding to the M5-brane corresponds
to $C_{\a+\b}$.
I don't know how to prove the statement about holomorphy but,
hopefully, it follows from the conjectured harmonicity
of $\ZM$ which we will discuss in the next section.

There is also another analogous situation for the type-IIA
little-string theory.
The type-IIA $U(N)$ little-string theory on $\MR{5,1}$ has
a moduli-space of $(\MS{1})^N/S_N$.
This moduli space is related to the compact scalars.
The radius of $\MS{1}$ is $M_s$ -- the scale of the little-string
theory. Now the ``center-of-mass'', i.e. the average of the $N$
scalars, is defined up to ${1\over N}M_s$. This means that after
compactification on $\MS{1}$, the theory will have topological sectors
according to the winding-number of this center-of-mass coordinate
around $\MS{1}$. This winding number is quantized in units of 
${1\over N}M_s$. If this $\MS{1}$ is in the $1^{st}$ direction, 
then $q$ units of this flux corresponds to having $q$ D4-branes
in directions $2\dots 6$.
On the other hand, if we take the $1^{st}$ direction to be time,
one can have, say,  $p$ units of flux that correspond to a charge
of $p$ D0-branes.
This flux means that the canonical momentum dual to the center-of-mass of
the compact scalars is ${{p N}\over {M_s}}$.
Now suppose we turn on a 1-form RR-flux $\sigma_1$
along the $1^{st}$ direction. In the notation of the previous sections,
this RR-flux corresponds to a root of $E_{8}$ as follows,
$$
e^{t_1 - t_7}\rightarrow \a'\equiv\a_1 +\a_2 +\a_3 +\a_4 +\a_5 +\a_6.
$$
The partition function of the sector with $p$ units of flux depends
on $\sigma_1$ via a factor $e^{i p \sigma_1}$.
On the other hand, in the D4-brane approach we use the RR-flux
$\sigma_{23456}$ of a D4-brane in directions $2\dots 6$.
This phase corresponds to the root,
$$
e^{t_2 +t_3 +t_4 +t_5 +t_6 +t_7}\rightarrow
   \a''\equiv \a_2 +2\a_3 +3\a_4 +4\a_5 + 3\a_6 + 2\a_7 + 2\a_8.
$$
As before, $\a'+\a''=\b'_0$ and the equivalence of the two approaches
to calculating the partition function in the given sector follows as
above.


\subsection{More on harmonic functions and instantons}\label{subsec:harm}
With the metric~(\ref{metna}), 
the volume form is proportional to,
$$
\sqrt{g} = e^{-2 \lrp{\lam}{\delta}},\qquad
\delta \equiv {1\over 2}\sum_{\a\in\Delta_{+}}\a.
$$
It can be shown that for the simple roots $\a_i$, $\delta$ satisfies,
$$
\lrp{\delta}{\a_i} = 1,\qquad i=1\dots r.
$$
The Laplacian can be written as,
$$
\nabla = {{\partial^2}\over {\partial\lam^2}}
   -2\lrp{\delta}{{\partial\over {\partial\lam}}}
  + 2\sum_{\a\in\Delta_{+}}
    e^{2\lrp{\lam}{\a}} {{\partial^2}\over {\partial C_{\a}^2}}
  +\cdots
$$
The ellipsis indicates terms cubic and quartic  in the $C_\a$'s and their
partial derivatives.
Let us denote the quadratic term by $\nabla_0$,
$$
\nabla_0\equiv
 {{\partial^2}\over {\partial\lam^2}}
   -2\lrp{\delta}{{\partial\over {\partial\lam}}}
  + 2\sum_{\a\in\Delta_{+}}
    e^{2\lrp{\lam}{\a}}{{\partial^2}\over {\partial C_{\a}^2}}.
$$
Let us also set $\nabla_1\equiv\nabla-\nabla_0$.
Pick a weight $\a$. The function,
$$
\Psi^{(\gamma)}_\a(\{C_{\b}\},\lam) =
e^{\lrp{\lam}{\gamma}
   -2\pi n e^{-\lrp{\lam}{\a}} + 2\pi i n C_\a},
$$
satisfies,
$$
\nabla_0\Psi =\left\{
4\pi^2 n^2 e^{-2\lrp{\lam}{\a}} (|\a|^2 - 2)
+4\pi n e^{-\lrp{\lam}{\a}} \lrp{\a}{\gamma-\delta-{\a\over 2}}
+\lrp{\gamma}{\gamma-2\delta}
\right\}\Psi.
$$
Thus, if $\a=\a_i$ is a simple root then we get
solutions for $\nabla_0\Psi_{\a_i} = 0$. The simplest is
to set  $\gamma=0$ and we also find 
another solution for $\gamma=2\delta$.
It can also easily be seen that $\nabla_1\Psi_{\a_i} = 0$.
Thus, we find the two solutions,
\bear
\Psi^{(0)}_{\a_i}(\{C_{\b}\},\lam) &=&
e^{-2\pi n e^{-\lrp{\lam}{\a_i}} + 2\pi i n C_{\a_i}},\nn\\
\Psi^{(2\delta)}_{\a_i}(\{C_{\b}\},\lam) &=&
e^{2\lrp{\lam}{\delta}
   -2\pi n e^{-\lrp{\lam}{\a_i}} + 2\pi i n C_{\a_i}},\nn
\eear
Now, for any real root $\a$,
we can find an element of the Weyl group $\Weyl$ that sends 
$C_{\a_i}$ to $C_{\a}$. If we apply it to $\Psi^{(0)}$ or
$\Psi^{(2\delta)}$ we obtain
the functions $\widetilde{\Psi}^{(0)}_\a$ and
$\widetilde{\Psi}^{(2\delta)}_{\a}$
that correspond to the instanton dual to the
phase $C_\a$.
In the limit $-\lrp{\lam}{\a}\rightarrow\infty$,
these functions behave, to leading order, as,
$$
\widetilde{\Psi}_\a(\{C_{\a}\},\lam) \sim
(\cdots) e^{-2\pi n e^{-\lrp{\lam}{\a}} + 2\pi i n C_\a},
$$

Now let us take two distinct simple roots $\a\equiv\a_i$ and
$\b\equiv\a_j$ such
that $\lrp{\a}{\b}=0$ and in particular,
$\a+\b$ is not a root (because $|\a+\b|^2 = 4$).
It can then be checked that,
\bear
\Psi^{(0)}_{\a_i,\a_j}(\{C_{\b}\},\lam) &\equiv&
e^{-2\pi n e^{-\lrp{\lam}{\a}} -2\pi m e^{-\lrp{\lam}{\b}}
 + 2\pi i n C_\a + 2\pi i m C_\b},\nn\\
\Psi^{(2\delta)}_{\a_i,\a_j}(\{C_{\b}\},\lam) &\equiv&
e^{2\lrp{\lam}{\delta}
   -2\pi n e^{-\lrp{\lam}{\a}} -2\pi m e^{-\lrp{\lam}{\b}}
 + 2\pi i n C_\a + 2\pi i m C_\b},\nn\\
&&\label{twoinst}
\eear
are also harmonic.
By applying the Weyl group (or U-duality) we obtain harmonic solutions
corresponding to instantons made by combining the instanton dual
to the phase $C_\a$ with the instanton dual to the phase $C_\b$.
As an example take $\a=\b_0$, the instanton dual to the M5-brane and
$\b=\b_1$, the instanton dual to the KK-monopole.
The functions~(\ref{twoinst}) describe an instanton made out of
the ${1\over 4}$-BPS state of an M5-brane together with a KK-monopole.

Now let us explicitly demonstrate these formulas for the example
of $SL(3,\BR)/SO(3)$. This example contains three positive
roots $\a,\b$ and $\a+\b$ which are all the ingredients that we need
so far.
It is convenient to write the Laplacian in the variables
of~(\ref{nilabl}) with the redefinition $b_0\equiv b - a c$ as,
\bear
\nabla &=&
{2\over 3}\px{\lambda}^2  -{2\over 3}\px{\lambda}\px{\mu}
+{2\over 3}\px{\mu}^2 -2\px{\lambda}
\nn\\ &&
+2 e^{2\lambda -2\mu}\px{a}^2
+2 e^{2\lambda -2\mu}(c^2 + e^{2\lambda + 4\mu})\px{b_0}^2
-4 e^{2\lambda -2\mu}c\px{a}\px{b_0}
+2e^{2\lambda + 4\mu}\px{c}^2.
\nn
\eear
The $SL(3,\BZ)$ transformation act as
$g\rightarrow g' = U g$. For the matrices
$$
U_a = \left(\begin{array}{rrr}
 1  &  1  &  0  \\
 0  &  1  &  0  \\
 0  &  0  &  1  \\
\end{array}\right),\,\,
U_b = \left(\begin{array}{rrr}
 1  &  0  &  1  \\
 0  &  1  &  0  \\
 0  &  0  &  1  \\
\end{array}\right),\,\,
U_c = \left(\begin{array}{rrr}
 1  &  0  &  0  \\
 0  &  1  &  1  \\
 0  &  0  &  1  \\
\end{array}\right),\,\,
$$
We have,
\bear
U_c &:& c\rightarrow c+1,\,\, b_0\rightarrow b_0-a,\nn\\
U_b &:& b_0\rightarrow b_0+1,\nn\\
U_a &:& a\rightarrow a+1,\nn
\eear
We will be looking for solutions that are invariant under
$U_a,U_b$ and $U_c$, but later we will also use the transformation,
\bear
a &\rightarrow& a' = -{{a}\over {e^{2\lam-2\mu} + a^2}},\nn\\
b &\rightarrow& b' = -c,\nn\\
c &\rightarrow& c' = b\nn\\
e^{\lam} &\rightarrow& e^{\lam'} =
  {{e^\mu}\over {\sqrt{1 + e^{2\mu-2\lam} a^2}}},\nn\\
e^{\mu} &\rightarrow& e^{\mu'} =
  e^{\lam}\sqrt{1 + e^{2\mu-2\lam} a^2}.\nn
\eear
generated by,
$$
U = \left(\begin{array}{rrr}
 0  &  -1  &  0  \\
 1  &  0  &  0  \\
 0  &  0  &  1  \\
\end{array}\right).
$$

First, we find the solutions,
\bear
\Psi^{(0)}_{\a} &=& e^{2\pi i n c -2\pi n e^{\lam+2\mu}},\nn\\
\Psi^{(\a+\b)}_{\a} &=& e^{3\lam + 2\pi i n c -2\pi n e^{\lam+2\mu}},\nn
\eear
Using the transformation of $U$ we can also generate,
\bear
\Psi^{(0)}_{n} &=&
 e^{2\pi i n b  -2\pi n e^{\mu+2\lam}\sqrt{1 + a^2 e^{2\mu-2\lam}}},\nn\\
\Psi^{(\a+\b)}_{n} &=&
(1 + a^2 e^{2\mu-2\lam})^{-3/2}
 e^{3\mu+ 2\pi i n b
   -2\pi n e^{\mu+2\lam}\sqrt{1 + a^2 e^{2\mu-2\lam}}},\nn
\eear
If we apply $U_a^{k/n}$ we find,
\bear
\Psi^{(0)}_{n,k} &=&
 e^{2\pi i n b  +2\pi i k c
    -2\pi \sqrt{n^2 e^{2\mu+4\lam} + (k+ n a)^2 e^{4\mu+2\lam}}},\nn\\
\Psi^{(\a+\b)}_{n,k} &=&
(1 + (a+{k\over n})^2 e^{2\mu-2\lam})^{-3/2}
 e^{3\mu+ 2\pi i n b +2\pi i k c
    -2\pi \sqrt{n^2 e^{2\mu+4\lam} + (k+ n a)^2 e^{4\mu+2\lam}}},\nn
\eear
In general, let $\a$ and $\b$ be two roots such
that $\gamma=\a-\b$ is also a root.
Let us pick one of the directions of $\MT{d}$ to be a Euclidean time.
Let us assume that the instanton that comes with the phase $e^{i C_\a}$
can be thought of as a BPS state with charge $Q_\a$ and a Euclidean
world line along that direction and let us assume that $e^{i C_\b}$
is associated with a similar instanton with charge $Q_\b$
and the same Euclidean world-line.
For example, $\a$ and $\b$ could correspond to two KK states that
come with an action of, say, $e^{t_1-t_2}$ and $e^{t_1-t_3}$, or any
of their U-dual descriptions (two M2-branes, two M5-branes, 
an M2-brane and a transverse KK momentum, and so on).

Under the assumptions, there is a BPS state with charge $k Q_\a + n Q_\b$
and it will come with a phase $e^{i n C_\a + i k C_\b}$.
The instanton action is,
$(n^2 e^{-2 \lrp{\lam}{\a}} + k^2 e^{-2 \lrp{\lam}{\b}})^{1/2}$.
We have seen in
subsection~(\ref{subsec:phxcst}) that we should set $C_\gamma = 0$.
Indeed, the function $\Psi^{(0)}_{n,k}$ found above behaves like,
\be\label{kpit}
e^{2\pi i n C_\a + 2 \pi i m C_\b 
  -2\pi\sqrt{n^2 e^{-2 \lrp{\lam}{\a}} + k^2 e^{-2 \lrp{\lam}{\b}}}},
\ee
when we substitute $b \equiv C_\a$, $c \equiv C_\b$ and set
$a \equiv C_\gamma = 0$.
A detailed analysis of the instanton terms of M-theory on $\MT{4}$ 
appears in \cite{KPI}.

To conclude this section,
we see that the harmonic equation has the potential of 
encoding all the BPS states
as well as states with less supersymmetry like 
the ${1\over 4}$-BPS states. We will later conjecture that the harmonic
equation actually encodes more than that.


\section{Some facts about infinite dimensional Kac-Moody algebras}
When we go to dimensions lower that 2+1D, the natural generalization
of the moduli-spaces are, formally, spaces built out of $E_9$ and 
$E_{10}$.

\parbox[c]{140mm}{
\begin{picture}(370,80)
%
%
\thicklines
\multiput(220,20)(20,0){8}{\circle{6}}
\multiput(223,20)(20,0){7}{\line(1,0){14}}

\put(320,23){\line(0,1){14}}
\put(320,40){\circle{6}}
\put(214,10){{\mbox{$\a_0$}}}
\put(236,10){{\mbox{$\a_1$}}}
\put(256,10){{\mbox{$\a_2$}}}
\put(276,10){{\mbox{$\a_3$}}}
\put(296,10){{\mbox{$\a_4$}}}
\put(316,10){{\mbox{$\a_5$}}}
\put(336,10){{\mbox{$\a_6$}}}
\put(356,10){{\mbox{$\a_7$}}}

\put(324,38){{\mbox{$\a_8$}}}
\put(220,40){{\mbox{$E_{9}$}}}
%
%
\thicklines
\multiput(20,20)(20,0){9}{\circle{6}}
\multiput(23,20)(20,0){8}{\line(1,0){14}}

\put(140,23){\line(0,1){14}}
\put(140,40){\circle{6}}
\put( 14,10){{\mbox{$\a_{-1}$}}}
\put( 36,10){{\mbox{$\a_0$}}}
\put( 56,10){{\mbox{$\a_1$}}}
\put( 76,10){{\mbox{$\a_2$}}}
\put( 96,10){{\mbox{$\a_3$}}}
\put(116,10){{\mbox{$\a_4$}}}
\put(136,10){{\mbox{$\a_5$}}}
\put(156,10){{\mbox{$\a_6$}}}
\put(176,10){{\mbox{$\a_7$}}}

\put(144,38){{\mbox{$\a_8$}}}
\put(20,40){{\mbox{$E_{10}$}}}

\end{picture}}

In the present paper, $E_9$ and $E_{10}$ are not related to
an extension of the notion of moduli-space. Nevertheless,
understanding the structure of these spaces is crucial
to the construction of the conjecture since they are related to
the \emph{domain} of the Master-function $\ZM$.

In this section we will review some facts about infinite dimensional
Kac-Moody algebras. Most of the information is taken from \cite{Kac}.
The infinite dimensional Lie algebras in which we will be interested
have simply-laced Dynkin diagrams so we will restrict to that case
as well.

\subsection{Definition}
The definition of the Lie algebra is as follows \cite{Kac}.
One starts with a finite symmetric Cartan matrix $A \equiv\{a_{ij}\}$
such that
all the diagonal elements $a_{ii} = 2$ and the nondiagonal elements
are negative integers. Here $i,j=1\dots r$ where $r$
is the rank ($=10$ for $E_{10}$).
One then picks independent vectors $\a_1,\dots,\a_r\in H=\MC{r+p}$
where $p$ is the dimension of the kernel of $A$. For $E_{10}$ we
have $p=0$ and for $E_9$ we have $p=1$.
One also defines the dual space $H^*$ which is also isomorphic
to $\MC{r+p}$
and picks vectors $\a^\vee_1,\dots,\a^\vee_r\in H^*$.
One defines the Kac-Moody Lie algebra $g(A)$ with
the generators $e_i,f_i$ ($i=1,\dots,r$) and $H$, by the following
relations (theorem 9.11 on p159 of \cite{Kac}):
\bear
 && \com{e_i}{f_j} = \delta_{ij}\a_i^\vee,\qquad
\com{h}{e_i} = \lrp{\a_i}{h} e_i,\qquad
\com{h}{f_i} = -\lrp{\a_i}{h} f_i,\nn\\
 && \com{h}{h'} = 0,\qquad
({\rm ad\ }e_i)^{1-a_{ij}} e_j = 0,\qquad
({\rm ad\ }f_i)^{1-a_{ij}} f_j = 0,\nn\\
&& (i,j=1,\dots,r,\,\, i\neq j\,\, h,h'\in H^*).\nn\\
&&\label{kmrel}
\eear

\subsection{Real and imaginary roots}\label{subsec:mthyrir}
There are several standard properties of finite dimensional Lie algebras
which do not go over to the infinite dimensional case.
We will now review some of these.

Let $\Delta\subset H$ be the set of roots and let $\Delta_+$
be the set of positive roots.
One defines the Weyl group $\Weyl$ as the group generated
by the $r$ reflections $\sigma_i$ of $H^*$ around the $r$ simple roots
$\a_i$. 
Also, one defines a bilinear form $(\cdot |\cdot)$
 on $H^*$ according to (p17 of \cite{Kac}),
$$
(\a_i |\a_j) = a_{ij}.
$$
Let us also denote the norm $|\a|^2 = (\a |\a)$ which is defined on
the real vector space $\oplus_{i=1}^r\BR\a_i$,
and let us define the lattice $Q=\oplus_{i=1}^r\BZ\a_i\subset H^*$
as the set of all points in $H^*$ that can be written as a linear
combination of the simple roots with integer coefficients.

Some facts about infinite dimensional Kac-Moody algebras
(which are not the case for finite Lie algebras) are (p55 of \cite{Kac}):

\begin{itemize}

\item
The bilinear form $(\cdot | \cdot)$ is not positive-definite.

\item
The Weyl group $\Weyl$ is infinite.

\item
The set of roots $\Delta$ is infinite.

\item
For every positive root $\a\in\Delta_{+}$ there exists $1\le i\le r$
such that $\a+\a_i$ is also a root.

\end{itemize}

A root $\a\in\Delta$ is called \emph{real}
if there exists $w\in\Weyl$ such
that $w(\a)$ is a simple root (p59 of \cite{Kac}).
A root $\a\in\Delta$ that is not real is called \emph{imaginary}.
Finite dimensional Lie algebras have no imaginary roots.
The properties of imaginary roots are very different from the real
roots.
Let us summarize some facts about imaginary roots from \cite{Kac}
(recall that we discuss only the simply-laced cases):

\begin{itemize}

\item
A root $\a$ is imaginary if and only if $(\a|\a)\le 0$
(p61 of \cite{Kac}).

\item
If $\a$ is an imaginary root and $p\in\BZ-\{0\}$
then $p\a$ is also an imaginary root (p63 of \cite{Kac}).

\item
For an imaginary root $\a$ there exists a unique root $\b$
which satisfies $\lrp{\b}{\a_i^\vee} \le 0$ for all $i=1,\dots,r$
and which is $\Weyl$-equivalent to $\a$ (p61 of \cite{Kac}).

\item
An imaginary root doesn't necessarily have to have a multiplicity of
one.

\end{itemize}

For affine and hyperbolic Kac-Moody algebras one can be even more 
specific.
A Kac-Moody algebra is called  \emph{affine} if its Cartan matrix $A$
is positive semidefinite of rank $r-1$ (p49 of \cite{Kac}).

A Kac-Moody algebra is called \emph{hyperbolic} if the Cartan matrix $A$
is of indefinite type and any connected proper subdiagram of the Dynkin
diagram is either finite or affine (p56 of \cite{Kac}).
The Kac-Moody algebra $E_{10}$ is an example.
For simply-laced affine Lie-algebras, $H^*$ has a one-dimensional
subspace of vectors with zero norm in the $(\cdot|\cdot)$
metric. The set of imaginary roots is
given by $n\delta_0$ for $n$ a nonzero integer and $\delta_0\in H^*$
is the minimal vector which satisfies $|\delta_0|^2 = 0$
(p64 of \cite{Kac}).
For a simply-laced hyperbolic Kac-Moody algebra the set of all imaginary
roots is equal to the set of all nonzero $\a\in Q\subset H^*$ with 
$|\a|^2\le 0$ (p67 of \cite{Kac}).
Also, for a finite, affine or hyperbolic Kac-Moody algebras,
the group of automorphisms of the lattice $Q\subset H^*$ which preserve
the bilinear form $(\cdot |\cdot)$ is a semidirect product of the 
group of automorphisms of the Dynkin diagram, parity transformation,
and the Weyl group $\Weyl$ (p68 of \cite{Kac}).

\subsection{$E_9$ as a current algebra}
Affine Lie-algebras can be realized as current-algebras.
We will describe this for $E_9$ (see \cite{Kac} for the generic case).
The Dynkin diagram of $E_9$ is obtained by adding an extra node to
the Dynkin diagram of $E_8$. Let $\a_0$ be the corresponding
simple root.
The root $\delta_0$ appearing in the characterization of
imaginary roots, in the subsection above, is equal to,
$$
\delta_0 = \a_0 + \theta,
$$
where,
$$
\theta = 2\a_1 + 3\a_2 + 4\a_3 + 5\a_4 + 6\a_5 + 4\a_6 + 2\a_7 + 3\a_8,
$$
is the root of $E_8$ with highest weight.
The generators of $E_9$ which correspond
to the positive real roots are given by the current-algebra form,
$$
J_0^{+\a},\,\, J_{+n}^{+\a},\,\, J_{+n}^{-\a},\qquad
\a\in\Delta^{(0)}_{+},\qquad n\in\BZ_{+},
$$
where $\Delta^{(0)}_{+}$ is the set of positive roots of $E_8$.
The corresponding weights are,
\be\label{affrts}
\a,\,\,\a + n\delta_0,\,\,-\a + n\delta_0.
\ee
The simple root $\a_0$ corresponds to the generator $J_{+1}^{-\theta}$
and the other simple roots correspond to the generators $J_0^{+\a_i}$.

The imaginary roots corresponding to $n\delta_0$ are 
$J_{+n}^a$ with $a=1\dots 8$ corresponding to the Cartan sublagebra
of $E_8$. The multiplicity of each of these roots is $r=8$.
The multiplicity of the real roots is 1.


\section{The physics of $E_9$ and $E_{10}$}
Now we will return to $E_9$ and $E_{10}$ and study the roots
in relation with the physics of M-theory on $\MT{9}$ and M-theory
and the twists.

\subsection{``M-theory on $\MT{9}$''}
Adding another node  $\a_0$ to the Dynkin diagram of $E_8$
corresponds, physically, to adding a $9^{th}$ radius, $R_9$ 
and adding a corresponding $s_0$ in (\ref{stot}) such that,
$$
s_0 = t_9 - t_1.
$$
The imaginary root $\delta_0$ would algebraically correspond
to an instanton with an action,
$$
V\equiv e^{t_1+\cdots+t_9} = {\rm Vol\ }(\MT{9}).
$$
Let us take a small detour to discuss
M-theory on $\MT{9}$. See also \cite{BFM} for related issues.
M-theory on $\MT{9}$ is a two-dimensional theory and its low-energy limit
is not a free theory. Thus, the usual
questions about the S-matrix and the Wilsonian effective action cannot
be asked.
On the other hand, $V^{-1}$ plays the role of a coupling constant.
Note that $V$ is invariant under U-duality and in the limit 
$V\rightarrow\infty$ the theory is described by a weakly coupled
$\sigma$-model \cite{SenTWOD}. Although $V$ is a fluctuating field,
it makes sense to ask about higher order corrections to the $\sigma$-model
action in the limit $V\rightarrow\infty$.\footnote{I have benefited from
discussions with S. Sethi on some of these issues and hope to explore
this subject further in collaboration with S. Sethi.}
The existence of real roots with weights $n\delta_0\pm\a$ (\ref{affrts})
 suggests the existence
of corrections that behave as $(\cdots)e^{-V^n}$.
(Recall from subsection~(\ref{subsec:harm}) that only roots with
norm 2 correspond to instantons and only the real roots have
norm 2.)

\subsection{``M-theory on $\MT{10}$''}
We will now list a few real roots of $E_{10}$ and their related
instantons.
The notation we use for the simple roots is $\a_1,\dots,\a_8$ for
the $E_{8(8)}$ generators and then $\a_0$ and $\a_{-1}$ for the two
added nodes.
It is actually more convenient to change the definition of the simple
roots such that $\a_{-1}$ will correspond to the phase that couples
to an instanton with action $e^{t_1-t_2}$.
%
It is convenient to make a change of bases in the root space $\MR{10}$
and define,
\be\label{vtchng}
\a_k \equiv \vt_{k+2} - \vt_{k+3},\,\, k=-1\dots 7,\qquad
\a_8 = \vt_8 + \vt_9 + \vt_{10}.
\ee
A root,
$$
\a = \sum_{k=1}^{10} m_k\vt_k,
$$
corresponds to an instanton with an action of $e^{\sum m_k t_k}$.
The metric in the new basis is,
$$
|\a|^2 = \sum_{k=1}^{10} m_k^2 - {1\over 9}\left(\sum_{k=1}^{10}m_k\right)^2.
$$
In this basis, the root lattice is given by all $\sum_{k=1}^{10} m_k\vt_k$,
such that $\sum_{k=1}^{10}m_k\in 3\BZ$ (i.e. divisible by 3).

Later on we will need the following three instantons.
The first is an M5-brane wrapping directions $1\dots 6$
that is given by $\b'_0$ as in~(\ref{defbzerp}), but since $\a_1\dots\a_8$
have changed we have to rewrite it in terms of the new roots,
\bear
\lefteqn{e^{t_1 + t_2 + t_3 + t_4 + t_5 + t_6} \rightarrow
\b_0 \equiv \vt_1+\vt_2+\vt_2+\vt_3+\vt_4+\vt_5+\vt_6
}\nn\\ &=&
\a_{-1} +2\a_0 +3\a_1 +4\a_2 +5\a_3 +6\a_4 +6\a_5 + 4\a_6 + 2\a_7 +2\a_8.
\label{defbzero}
\eear
A KK-monopole with respect to the $8^{th}$ circle and wrapping directions
$1\dots 7$ is given by $\b'_1$ as defined in~(\ref{defbonp}).
In the new variables it is,
\bear
\lefteqn{
e^{2 t_8 + \sum_1^7 t_j} \rightarrow
\b_1\equiv 2 \vt_8 + \sum_1^7 \vt_j
}\nn\\ &=&
\a_{-1} +2\a_0 +3\a_1 +4\a_2 +5\a_3 +6\a_4 +7\a_5 + 6\a_6 + 3\a_7 +3\a_8.
\label{defbone}
\eear
If we reduce along the $10^{th}$ direction to get
type-IIA, we can write the instanton which formally
corresponds to a D8-brane wrapping directions $1\dots 9$,
\bear
\lefteqn{
e^{3 t_{10} + \sum_1^9 t_j} \rightarrow
\b_2\equiv 3\vt_{10} + \sum_1^9\vt_j
}\nn\\ &=&
\a_{-1} +2\a_0 +3\a_1 +4\a_2 +5\a_3 +6\a_4 +7\a_5 + 4\a_6 + \a_7 +4\a_8.
\label{defbtwo}
\eear

Since all these instantons are U-dual to a KK-particle,
and since U-duality is related to the Weyl group $\Weyl$, as we explained
above, it follows that all these roots are \emph{real} roots.
We calculate,
\be\label{bbprods}
|\b_1|^2 = |\b_2|^2 = |\b_0|^2 = 2,\qquad
(\b_0,\b_1)=0,\,\,
(\b_0,\b_2)=-2,\,\,
(\b_1,\b_2)=-3.
\ee
We will also need the roots which correspond to the Dehn twists
of the $8^{th}$ direction as one goes around the $j^{th}$ direction
for $j=1\dots 6$.
This root also corresponds to an instanton made of a KK-particle
along the $8^{th}$ direction with a world-line along the $j^{th}$.
They are,
\be\label{defgam}
\gamma_j\equiv \vt_j-\vt_8 = \sum_{k=j-2}^5 \a_k,\qquad j=1\dots 6
\ee
Similarly, we take the roots which correspond to Dehn twists
of the $10^{th}$ direction as one goes around the $j^{th}$ direction,
\be\label{defgamp}
\gamma'_j\equiv \vt_j-\vt_{10} =
\sum_{k=j-2}^7 \a_k,\qquad j=1\dots 6.
\ee
These roots satisfy,
\be\label{gamgam}
(\gamma_i,\gamma_j) =
(\gamma'_i,\gamma'_j) = 1 + \delta_{ij},\qquad
(\gamma_i,\gamma'_j) = -\delta_{ij}.
\ee
\bear
&&
 (\b_0,\gamma_j) = 1,\,\,
 (\b_1,\gamma_j) = -1,\,\,
 (\b_2,\gamma_j) = 0,\nn\\
&&
 (\b_0,\gamma'_j) = -1,\,\,
 (\b_1,\gamma'_j) = -1,\,\,
 (\b_2,\gamma'_j) = 2.\label{bgprod}
\eear
Note that $\b_2-\gamma_j'$ is a combination of simple roots
with positive integer coefficients. Since $|\b_2 - \gamma_j'|^2 = 0$,
it follows from the characterization of the roots of hyperbolic
Kac-Moody algebras in subsection~(\ref{subsec:mthyrir}) that
$\b_2-\gamma_j'$ is an imaginary root.
Moreover, $\b_2-2\gamma_j'$ is also a root and in fact corresponds
to a D8-brane when the $j^{th}$ direction is being used to reduce from
M-theory to type-IIA.
Similarly, we check that $|\b_2 + \b_0|^2 = 0$ and 
$|\b_2 + \b_1| = -2$. Thus, $\b_2+\b_0$ and $\b_2 + \b_1$ are imaginary
roots.


\section{The conjecture}
Now we are ready to formulate the conjecture mathematically.
The physical idea is that we take the partition function $\ZM$
which is a function of $E_{10}(\BZ)\backslash E_{10}(\BR)/\Kom$.
We replace $E_{10}(\BR)/\Kom$ with the probably equivalent
$\Nil\Abl$ decomposition into a product of a nilpotent and an
abelian element.
The Group $\Abl$ is identified with $SO(1,1)^{10}$ and 
we denote the element of $\Abl$ as,
$$
{\mbox{diag}}(e^{t_1}, e^{t_2},\cdots,e^{t_{10}}).
$$
We wish to extract out of $\ZM$ the twisted partition function
of the little-string theory of $N$ NS5-branes compactified on
$\MT{6}$ with radii $r_1,\dots,r_6$.
We will also take $R_7 = \lam_{IIA}^{2/3}$ to be the radius of a circle
for reduction of M-theory to type-IIA.
We first identify,
$$
t_k \equiv \log R_k = \log (\lam^{-1/3} r_k),\,\, k=1\dots 6,\qquad
t_7 \equiv \log R_7 = \log (\lam^{2/3}).
$$
Let us assume that the $SU(2)\times SU(2)$ twists are given by
$$
(e^{i\omega_k},e^{i\omega'_k})\in SU(2)\times SU(2),\qquad k=1\dots 6.
$$
It is convenient to think of an auxiliary $\MT{10}$ with sides
of lengths $t_j=\log R_j$ ($j=1\dots 10$).
Nevertheless, we stress that M-theory on $\MT{10}$ is not a well-defined
notion (at least it is not clear to me what are the calculable quantities
for M-theory on $\MT{10}$).
We wish to identify $\omega_k$ as Dehn twists of the $8^{th}$ circle
(of radius $R_8$) around the $k^{th}$ circle. Similarly, we wish to
identify ${1\over 2}(\omega_k + \omega'_k)$
(the reason why this combination
and not just $\omega'_k$ appears here will be explained shortly)
with Dehn twists of the $10^{th}$ circle around the $k^{th}$ circle
($k=1\dots 6$).

As we explained in section~\ref{subsec:zmhdt} (following
\cite{CGKM}) the identification of $\omega_k$ with the Dehn twist
of the $8^{th}$ direction can be achieved by immersing the M5-brane
inside a KK-monopole with respect to the $8^{th}$ direction.
We need a similar construction that will allow us to identify
${1\over 2}(\omega_k + \omega'_k)$ with the Dehn twists of, say, the
$10^{th}$ circle.
It seems that the natural thing to try is to immerse the construction,
that we have so far, inside a ``bigger'' brane.
In this case, we will take what would be called a D8-brane if 
M-theory were to be reduced to type-IIA along the $10^{th}$ direction.
The corresponding instantonic object was described in equation
(\ref{defbtwo}).
I do not have a ``geometrical-proof'' for this construction and the
conjecture is based on the fact that algebraically this looks analogous
to the way we realized the twist $\omega_k$ with the help of the
KK-monopole.
A ``geometrical-proof'' is harder to construct because we are not 
interpreting $E_{10}$ as the ``moduli-space'' of M-theory on 
$\MT{10}$.
Moreover, a D8-brane of type-IIA is not an object that can exist
on its own. It needs orientifold planes to stop the dilaton
from exploding \cite{PolWit}. Having said that, 
one could still argue, heuristically, as follows.
Recall that a D8-brane changes the gradient of $1/\lam_s$.
Suppose we start with a situation for which $\lam_s=\lam_0$ is
constant to the left of the D8-brane and then increases to the
right of the D8-brane.
As explained in \cite{PolWit} the dilaton will reach a singularity
after a finite distance from the D8-brane.
Suppose that we try to push this distance as far away from the D8-brane
as we can. We can try to take the limit $\lam_0\rightarrow 0$.
Now let us go back to M-theory.
$\lam_s$ is related to the radius of the $10^{th}$ circle and in the
heuristic construction above, it seems that this radius is zero
at the position of the D8-brane and increases away from it to the
right. Furthermore, rotations of the $10^{th}$ circle indeed seem
to be identified with the transverse $SO(2)$ rotations corresponding
to ${1\over 2}(\omega_k + \omega'_k)$.
We will, however,
 not put much emphasis on the geometrical argument and stress that
our motivation is algebraic.
(The argument given above does not quite agree with the lifting
of the exact metric derived in \cite{PolWit} to 11D.)

The final step is to extract from $\ZM$ the contribution of 
an instanton made by (formally) wrapping $N$ M5-branes on directions
$1\dots 6$ and adding to it a KK-monopole with respect to
the $8^{th}$ circle wrapping directions $1\dots 7$ and finally a
D8-brane of type-IIA wrapping directions $1\dots 6,8,9$.
The contribution of this instanton is isolated by Fourier transforming
with respect to the appropriate phases.
We will explain in detail how to treat the various phases below.
Finally we take the limit,
$$
t_6,t_8,t_9\rightarrow\infty
$$
and $t_7\rightarrow 0$ keeping $r_1\dots r_6$ fixed as in \cite{SeiVBR}.
At the same time, we must decide what limit to take for $t_{10}$.
We will address that issue later.
The instanton contribution will have the trivial factor
because of the tensions of the M5-branes, KK-monopole and D8-brane.
Naively, this factor looks like the sum of the tensions,
$$
{\mbox{exp}}\{-N e^{t_1+\cdots+t_6} -e^{2t_8 + t_1+\cdots +t_7}
          -e^{3t_{10} + t_1+\cdots +t_9}\}.
$$
However, since $\lrp{\b_2}{\b_1}\neq 0$ and $\lrp{\b_2}{\b_0}\neq 0$
this is not an approximate harmonic function and this factor
will have to be modified. We will return to that point in
subsection~(\ref{subsec:tten}).
The conjecture is that after we clean the (modified)
factor, we get the desired
partition function of the twisted little-string theory.
This partition function
encodes, among other things, the spectrum and S-matrix of ordinary
non-supersymmetric Yang-Mills theory as explained above.

The reason for defining the Dehn twists of the $10^{th}$ circle
to be ${1\over 2}(\omega'_k +\omega_k)$ and not just $\omega'_k$
is that, as we interpreted $\ZM$ in the previous sections,
The $SO(2)$ which generates  translations of the $10^{th}$ direction
is interpreted as embedding $SO(2)$ in the diagonal $SU(2)$ of
$SU(2)\times SU(2)=Spin(4)$ such that the vector $\rep{4}$ of $Spin(4)$
decomposes as $\rep{2}+\rep{1}+\rep{1}$. This is because this $SO(2)$
was interpreted as rotations of 2 out of the 4 dimensions transverse
to the NS5-brane.

\subsection{The phases}\label{subsec:conjph}
What do we do with the other $C_{\a}$ phases that do not
participate in the construction?
As we have seen in section (\ref{subsec:phxcst}),
if $\ZM$ is periodic in $C_{\a}$ and $C_{\b}$ for some $\a$ and
$\b$ corresponding to generators $\tau^{+\a}$ and
$\tau^{+\b}$
then $e^{i C_{\a+\b}}$ that corresponds to 
$\tau^{+(\a+\b)}\equiv\com{\tau^{+\a}}{\tau^{+\b}}$ is a section
of a nontrivial line-bundle over the torus parameterized
by $(C_{\a},C_{\b})$.
To extract a single-valued function we can do one of two things:

\begin{itemize}
\item
We can integrate over the fiber $C_{\a+\b}$ to get a periodic function
on the base parameterized by $(C_{\a},C_{\b})$ or,
\item
We can set $C_{\b}=0$ to get a single-valued function on
the torus parameterized by $(C_{\a},C_{\a+\b})$.
\end{itemize}

In the procedure outlined above there are phases $C_{\a}$ that
we have to keep periodic. These are the $\omega_k$ and
$\omega_k'$ twists, the Dehn twists
of $\MT{6}$, and
the phases dual to the M5-brane, KK-monopole and D8-brane.
Let $L_1\subset\Lambda_{+}$ denote this finite set of roots that 
we need to keep periodic.
The prescription for the other phases will be,

\begin{itemize}
\item
The phases corresponding to roots in $L_1$ are reserved for
``special treatment.''
\item
If $\b$ is a positive root that can be written as $\b=\g-\a$
for $\a,\g\in L_1$, then $C_{\b}$ should be set to zero.
This procedure should be repeated inductively. Thus,
$C_{\b_0-2\gamma_k}$, for example, is also set to zero.
\item
All other phases should be integrated.
\end{itemize}
Note that in the infinite dimensional Lie-algebra there are some
roots that appear with a multiplicity higher than $1$.
We have been using the notation $C_{\a}$ rather sloppily because
in such a case there are more than one $C_{\a}$ but it is easy
to extend the procedure to such cases.

Let us denote by $\BTW$ the space of $t_1\dots,t_{10}$ together
with the phases in $L_1$ and the phases that are set to $0$.
$\BTW$ is a finite dimensional manifold.

Now let $C_{\a}$ and $C_{\b}$ be two phases such that
we have to integrate over $C_{\a+\b}$.
If now $C_{\a+2\b}$ is also a root then we should integrate over
it as well.  In the case of an infinite Lie-algebra this
procedure could continue ad infinitum but, for generic points 
in $\MTW$, the size of the circles corresponding to a root $C_{\gamma}$
decreases exponentially as we go to higher $\gamma$'s 
and the process could be convergent.

There is actually an advantage in integrating over all but a finite
number of phases. Later on we will conjecture that $\ZM$ is harmonic.
Recall from subsection~(\ref{subsec:harm})
 that some of the approximate solutions to the Laplace
equation contain a factor of $e^{\lrp{\lam}{\delta}}$ where
$\delta$ is the sum of all the positive roots. For an infinite dimensional
Lie-algebra such a factor will diverge. However, if we integrate on
an infinite number of phases we are going to get back a contribution
of $e^{-\lrp{\lam}{\a}}$ for each $C_{\a}$ that we integrate
(this is just the size of the circle) and the divergence will cancel.

With $\b_0,\b_1,\b_2,\gamma_k,\gamma'_k$ defined in 
(\ref{defbzero},\ref{defbone},\ref{defbtwo},\ref{defgam},\ref{defgamp}),
let us denote the periodic phases,
$$
\phi \equiv C_{\b_0},\,\,
\psi \equiv C_{\b_1},\,\,
\chi \equiv C_{\b_2},
$$
Let us also set,
$$
\omega_k = C_{\gamma_k},\qquad
{1\over 2}(\omega_k +\omega'_k) = C_{\gamma'_k}.
$$

\subsection{The variable $t_{10}$}\label{subsec:tten}
We have to decide what to do with the variable $t_{10}$.
The two possibilities to consider seem to be either
$t_{10}\rightarrow\infty$ or $t_{10}\rightarrow -\infty$.
I do not have a good argument for this, but it seems that the
correct thing to do is to take $t_{10}\rightarrow \infty$.
This is analogous to taking $t_8\rightarrow\infty$ to 
decouple the KK-monopole.
On top of that, the erratic behavior of $\ZM_9$ or $\ZM_7$
alluded to in section (\ref{erratic}) could be attributed
to the limit $t_{10}\rightarrow\infty$.
Thus, we conjecture that $\ZM$ is well-behaved
but when $\ZM_9$ is extracted from $\ZM$ we need to Fourier transform
with respect to an appropriate phase and then
multiply by the factor $e^{3t_{10} + t_1 + \dots + t_9}$
which blows up
in the limit $t_{10}\rightarrow \infty$.

\subsection{Conjecture 1}
We propose that there exists a unique function $\ZM$ defined on
$E_{10}(\BZ)\backslash \Nil\Abl$ such that the partition
function 
$Z_{N,k}(r_1,\dots,r_6,\omega_1,\dots,\omega_6,\omega'_1,\dots,\omega'_6)$
of the little-string theories of $N$ NS5-branes at an $A_{k-1}$ singularity
\cite{IntNEW} with $Spin(4)_R$-symmetry twists
$e^{i\omega_j}\otimes e^{i\omega'_j}\in SU(2)\otimes SU(2) = Spin(4)$
($j=1\dots 6$) can be calculated from $\ZM$ as,

\bear
\lefteqn{
Z_{N,k}(r_1,\dots,r_6,\omega_1,\dots,\omega_6,\omega'_1,\dots,\omega'_6) =}
\nn\\
 && e^{N e^{t_1+\cdots+t_6}}
\lim_{t_7\rightarrow -\infty} \lim_{t_8,t_9\rightarrow\infty}
\lim_{t_{10}\rightarrow \infty}
 e^{k e^{2t_8 + t_1+\cdots +t_7} +e^{3t_{10} + t_1+\cdots +t_9}}
\nn\\ &&
\int\cdots\int d\phi d\psi d\chi\, e^{i N \phi + i k \psi + i\chi}
\log\ZM(t_1 = \log R_1 - {1\over 2}t_7,\dots,
 t_6 = \log R_6 - {1\over 2}t_7,\nn\\
&&
t_7,\dots, t_{10},\,
\phi,\psi,\chi,\omega_1,\dots,\omega_6,\omega'_1\dots\omega'_6,
\cdots).\label{conjone}
\eear

This formula extracts the contribution of the M5+KK+D8 instanton
we discussed above.
The $\int\cdots$ and the ellipsis in $\log\ZM$ denotes what we do with the
other $C_{\a}$'s. As we explained above we set some of 
them to zero and we integrate over others.

In particular, for $k=1$ we conjecture that we get the twisted 
partition functions of the little-string theories of \cite{SeiVBR}.

\subsection{Conjecture 2}
Based on the fact that the leading contributions (i.e. the brane
tension and phase) to $\ZM$ seem to be harmonic we conjecture that
$\ZM$ is a unique harmonic function defined on
$E_{10}(\BZ)\backslash\Nil\Abl$ which goes to zero in the various classical
limits.
This conjecture is also based on the bias that this
seems to be the simplest equation that $\ZM$ could
satisfy!

Note that if we integrate $\ZM$ over the phases
as in the prescription in subsection~\ref{subsec:conjph} we are left
with a harmonic function defined over the finite dimensional manifold
$\BTW$ (see subsection~\ref{subsec:conjph}).
The prescription above was to actually integrate $\log\ZM$ and 
not $\ZM$ itself. Nevertheless it would be interesting to see how
harmonic functions over $\BTW$ behave.
Since $\BTW$ contains $t_{10}$ it could be that the boundary conditions
are not specified uniquely and it is only when we include all the
other infinite number of phases and use the restrictions of $E_{10}(\BZ)$
we get well-defined boundary conditions.
In this context, we would like to mention a possibly related idea
discussed in \cite{BanSus} and a proposal
of E. Witten about $E_9(\BZ)$ (see footnote in \cite{BanSus}).\footnote{I
am grateful to E. Witten for telling me about this.}
The proposal was that the U-duality group $E_9(\BZ)$ for M-theory
on $\MT{9}$ which contains elements that exchange canonical coordinates
with canonical momenta (recall that in 2+1D the U-duality group contains
elements which transform scalars to duals of vectors and therefore
in 1+1D there are elements which transform scalars to their duals)
might be strong enough to ensure that there is only a single wave function
that satisfies all the constraints!


\section{Discussion}
We have proposed two conjectures.
The first is that there exists a well defined function over a coset of
the group manifold of $E_{10}$ which encodes all the partition functions
of gauge-theories, $(2,0)$ and little-string theories.
The second conjecture was that this function satisfies Laplace's
equation with respect to the $E_{10}$-invariant metric.

We would like to conclude with a few suggestions for extensions
of the conjectures:

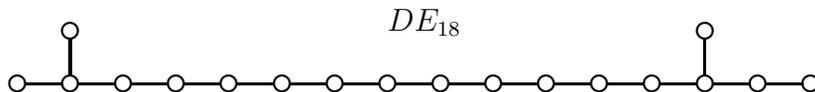
\parbox[c]{140mm}{
\begin{picture}(400,80)
\thicklines
\multiput(20,20)(20,0){16}{\circle{6}}
\multiput(23,20)(20,0){15}{\line(1,0){14}}

\put(280,23){\line(0,1){14}}
\put(280,40){\circle{6}}

\put(40,23){\line(0,1){14}}
\put(40,40){\circle{6}}

\put(160,40){{\mbox{$DE_{18}$}}}
\end{picture}}\newline

\begin{itemize}

\item {\em Heterotic string theories and $DE_{18}$}

A natural
generalization of the construction to the heterotic string theories
on $\MT{9}$ with twists seems to be to replace $E_{10}$ with
the Kac-Moody algebra based on the Dynkin diagram $DE_{18}$
in the picture (see p57 of \cite{Kac} for the notation).
This Dynkin digram is not hyperbolic so the characterization
of the root lattice in subsection~(\ref{subsec:mthyrir}) could be
more complicated.
The conjecture is that
a harmonic function on the coset of $DE_{18}$ will encode the
partition functions of the $(1,0)$ little-string theories \cite{SeiVBR}
and the $(1,0)$ 5+1D CFTs with global $E_8$ symmetry on $\MT{6}$.
It will also encode the $(2,0)$ and little-string theories on
$K_3\times \MT{2}$.

\item {\em Minkowski metric}

In the discussion in previous sections we have not been quite specific
as to whether the space-time metric is Euclidean or Minkowskian.
In principle, M-theory requires a Minkowskian metric because it contains
real 10+1D fermions.
In our setting we should take the time direction to be inside the
compactified $\MT{9}$.
In general, if $l$ of the internal directions of $\MT{d}$ are 
time-like, the moduli space is replaced by (see a recent discussion
in \cite{Minkow}),
$$
E_{d(d)}(\BZ)\backslash E_{d(d)}(\BR)/K',
$$
where $K'$ is a noncompact version of the compact group $K$.
For example, 
\be
\begin{array}{ccc}
d=4 & K = SO(4) & K' = SO(4-l, l) \\
d=5 & K = SO(5)\times SO(5) & K' = SO(5-l, l)\times SO(5-l,l) \\
\cdots & & \\
\end{array}
\ee

\item {\em Calculable partition functions in general}

More generally, we would like to ask whether harmonic functions on
other infinite dimensional Lie algebras
(and in particular those of Kac-Moody  type) encode information
about other vacua of M-theory.
Related to that is the question: {\em ``what are the quantitative
questions in M-theory?''}
In general we can ask about the partition function given some boundary
conditions. The question about the S-matrix of gravitons is of this
nature and the AdS/CFT correspondence \cite{Juan} with the interpretation
of \cite{GKP,WitAdS} is, of course, the most concrete and known example.
In fact the questions about the S-matrix can be derived from the AdS/CFT
correspondence \cite{PolAdS}.
The questions we were asking in this paper are also of this nature
because the twists can be defined as the boundary conditions at infinity.
We would also like to mention another example of a quantitative
question which encodes the answer to field theory questions.
This is the correspondence found in 
\cite{GopVafI}-\cite{GopVafIV}
between Chern-Simons
theories and string theory partition functions.

\end{itemize}

Let us also suggest a few topics for further research:

\begin{itemize}

\item {\em The U-duals of twists}

Independently of whether conjecture (2), or even conjecture (1),
is right or wrong, we have seen that M-theory on $\MT{7}$ contains
interesting types of $Spin(4)$ twists which are U-dual to the 
geometrical ones.
In particular we would like to recall the twist by the KK-monopole
charge which might have an interpretation in terms of some
generalized geometry.

\item {\em The spectrum of the little-string theories}

The question of whether the little-string theories are well-defined
or not has been the subject of some debate recently
(see \cite{MalStr,ABKS}).
In this paper, we were operating under the assumption that
the partition function of the little-string theory on $\MT{6}$
with $Spin(4)$ twists is a well-defined function. However, unlike
the $(2,0)$ theory on $\MT{5}$ (with twists),
it seems that the little-string theories
on $\MT{5}$ do not have a discrete spectrum.
This interpretation seems to be consistent with the results of
\cite{GubKle}.
Thus, the partition function on $\MT{6}$ could be well-defined but
not have a discrete Fourier transform with respect to time.

\item {\em Electric/magnetic fluxes and confinement}

If we had the partition functions $Z_N$ for generic twists,
we could check confinement by checking the behavior in the sector
with electric flux.
For example, the $(2,0)$ theory compactified on $\MS{1}\times\MS{1}$
with radii $R_5\ll R_6$ and with a generic twist along $R_6$
(and zero twists along the other directions)
can be approximated by ordinary Yang-Mills theory at energy scales below
$R_6^{-1}$ (according to the conjecture in \cite{WitAdSII}).
 At that scale the coupling constant will be of the order of,
$$
-i\tau = {{4\pi}\over {g^2}} = {{R_6}\over {R_5}}.
$$
It will then flow with the $\beta$-function of Yang-Mills theory,
$$
\beta(g) = -{{g^3}\over {(4\pi)^2}} {{11 N}\over 3},
$$
and so the QCD scale will be,
$$
\Lambda \sim R_6^{-1} e^{-{{6\pi R_6}\over {11 N R_5}}}.
$$
Thus, the partition function in the sector with one unit of electric
flux along $R_1$ (with $R_0$ taken as time)
should behave as (see \cite{tHfFLX}),
$$
Z_N\sim {\mbox{\rm exp}}\left\{
-f(\omega_6,\omega_6') {{R_1 R_0}\over {R_6^2}}
 e^{-{{12\pi R_6}\over {11 N R_5}}}
\right\},
$$
when,
$$
R_1,R_0\gg R_6 e^{{{6\pi R_6}\over {11 N R_5}}},\qquad
R_6\gg R_5.
$$
Here $f(\omega_6,\omega_6')$ is an unknown function of the twists.
It determines the prefactor in $\Lambda$.
It would be interesting to see if this function arises as a limit of
a harmonic function on $E_{10}$.

\item {\em Gravity}

Can we get information about gravity from $E_{10}$?
Although I do not have a precise prescription, it seems that 
$Z_0$ in (\ref{zphi}) should encode the information about
the S-matrix of gravitons. If conjecture (1) is true,
it would be interesting to make this more precise.

\item {\em Definition of the Laplacian on $E_{10}$}

Finally, there might be some subtleties involved
in the definition of the Laplacian
on the infinite dimensional coset
$\MTW=E_{10}(\BZ)\backslash E_{10}(\BR)/\Kom$.
Although this is not the first time infinite dimensional manifolds
are encountered in physics, and even the kinetic part of the Hamiltonian
of a field-theory is a Laplacian on the infinite dimensional 
configuration space, we need to have a precise definition.
In our case there are two facts that could be important for the definition.
As we have seen, one can separate the variables of $\MTW$
into 10 noncompact ones and the rest (which are infinite in number)
are compact circles. Each circle corresponds to a positive 
root of the Lie algebra. For a generic fixed value of the
10 noncompact variables, and given an arbitrarily small
$\Lambda$, the number of circles with radius
bigger than $\Lambda$ is finite.

\end{itemize}


\section*{Acknowledgments}
I have benefited from various discussions with
M. Krogh, N. Seiberg, S. Sethi and E. Witten
on issues related to this paper.
I am also grateful to O. deWolfe, T. Hauer, L. Motl,
H. Nicolai and B. Pioline for helpful correspondence.


\def\np#1#2#3{{\it Nucl.\ Phys.} {\bf B#1} (#2) #3}
\def\pl#1#2#3{{\it Phys.\ Lett.} {\bf B#1} (#2) #3}
\def\physrev#1#2#3{{\it Phys.\ Rev.\ Lett.} {\bf #1} (#2) #3}
\def\prd#1#2#3{{\it Phys.\ Rev.} {\bf D#1} (#2) #3}
\def\ap#1#2#3{{\it Ann.\ Phys.} {\bf #1} (#2) #3}
\def\ppt#1#2#3{{\it Phys.\ Rep.} {\bf #1} (#2) #3}
\def\rmp#1#2#3{{\it Rev.\ Mod.\ Phys.} {\bf #1} (#2) #3}
\def\cmp#1#2#3{{\it Comm.\ Math.\ Phys.} {\bf #1} (#2) #3}
\def\mpla#1#2#3{{\it Mod.\ Phys.\ Lett.} {\bf #1} (#2) #3}
\def\jhep#1#2#3{{\it JHEP.} {\bf #1} (#2) #3}
\def\atmp#1#2#3{{\it Adv.\ Theor.\ Math.\ Phys.} {\bf #1} (#2) #3}
\def\jgp#1#2#3{{\it J.\ Geom.\ Phys.} {\bf #1} (#2) #3}
\def\hepth#1{{\it hep-th/{#1}}}

\end{document}